\documentclass[lettersize,journal]{IEEEtran}
\usepackage{amsmath,amsfonts}
\usepackage{algorithmic}
\usepackage{algorithm}
\usepackage{array}
\usepackage{textcomp}
\usepackage{stfloats}
\usepackage{url}
\usepackage{verbatim}
\usepackage{graphicx}
\usepackage{color}
\usepackage{multirow}
\usepackage{subcaption}
\usepackage{booktabs}
\usepackage{bbding}
\usepackage{cite}
\begin{document}

\title{Learned Video Compression via Heterogeneous Deformable Compensation Network}

	\author{Huairui Wang, \emph{Student Member, IEEE}, Zhenzhong Chen\textsuperscript{*\thanks{This work was supported in part by National Natural Science Foundation of China (Grant No. 62036005) and Hong Kong Research Grants Council (GRF-15213322). (* Corresponding author: Zhenzhong Chen.)}\thanks{Huairui Wang and Zhenzhong Chen are with the School of Remote Sensing and Information Engineering, Wuhan University, Wuhan 430079, China (e-mail: wanghr827@whu.edu.cn; zzchen@ieee.org).}\thanks{Chang Wen Chen is with The Hong Kong Polytechnic University, Hong Kong, China (e-mail: changwen.chen@polyu.edu.hk).}
	}, \emph{Senior Member, IEEE}, \\and Chang Wen Chen, \emph{Fellow, IEEE}
}


\maketitle

\begin{abstract}
Learned video compression has recently emerged as an essential research topic in developing advanced video compression technologies, where motion compensation is considered one of the most challenging issues. In this paper, we propose a learned video compression framework via heterogeneous deformable compensation strategy (HDCVC) to tackle the problems of unstable compression performance caused by single-size deformable kernels in downsampled feature domain. More specifically, instead of utilizing optical flow warping or single-size-kernel deformable alignment, the proposed algorithm extracts features from the two adjacent frames to estimate content-adaptive heterogeneous deformable (HetDeform) kernel offsets. Then we align the features extracted from the reference frames with the HetDeform convolution to accomplish motion compensation. Moreover, we design a Spatial-Neighborhood-Conditioned Divisive Normalization (SNCDN) to reduce spatial statistic dependencies and achieve more effective data Gaussianization combined with the Generalized Divisive Normalization. Furthermore, we propose a multi-frame enhanced reconstruction module for exploiting context and temporal information for final quality enhancement. Experimental results indicate that HDCVC achieves superior performance than the recent state-of-the-art learned video compression approaches.
\end{abstract}


\section{Introduction}

The past few years have witnessed an explosion of video content on the Internet, bringing significant challenges in network bandwidth and data storage. Therefore, developing more efficient video compression schemes has attracted substantial attention in the video coding community. 

Traditional video compression methods, such as H.264\cite{h264} and HEVC\cite{HEVC}, have been designed in a hybrid style that heavily relies on hand-crafted techniques to improve compression efficiency. With block motion estimation, discrete cosine transform (DCT), entropy coding, and other related technologies, these conventional video compression methods have achieved stable performance in exploiting data redundancy. Based on the general framework of these traditional methods, several deep neural networks have been designed into the hybrid framework as sub-modules\cite{9399291,9185043}. These kinds of solutions aim to improve the performance of certain particular modules of the whole compression framework. Although each sub-module is well designed or improved by DNN individually, the entire framework cannot be optimized in an end-to-end fashion. It is desirable to further enhance video compression performance by jointly optimizing the whole framework.

In order to optimize the compression framework in an end-to-end fashion, some attempts have been made in the previous work\cite{Wu:2018,Lu2019CVPR,yang2020learning}. More recently, several solutions\cite{Lu2019CVPR,rippel2019learned,liu2020learned} adopt optical flow as motion information to achieve an end-to-end training. These algorithms exploit the possibility of replacing the block-based motion vectors in the traditional video coding by optical flow field. The decompression quality and bits consumption shall depend on the optical flow prediction accuracy and warping operation efficiency. Nevertheless, the flow-based compression methods usually rely on a complex pre-trained model to predict the optical flow. If not, inaccurate flow estimation may result in weird reconstruction artifact during the pixel domain warping\cite{tian2020tdan}.

Deformable convolution has been applied in the previous work\cite{hu2021fvc} for feature alignment, achieving more robust and effective motion compensation. By estimating kernel offset maps, deformable convolution can be integrated into the feature space to achieve desired compensation. It is worth noting that FVC\cite{hu2021fvc} performs deformable convolution on the features with the exact resolution as the original image. We carry out deformable convolution in downsampled feature domain to avoid the large GPU memory consumption caused by deformable convolution on the original resolution and accelerate the model inference speed. However, from our exploration experiments, we derive that deformable-compensation-based methods with single kernel sizes have limitations in certain bit rates in downsampled feature domain. In other words, single-size deformable kernels cannot be a globally optimal choice for different target bit rates in this condition. Besides, according to our exploration experiments in Section \ref{sec:HetDeform Compensation}, the rate-distortion performance of deformable-compensation-based methods is not positively correlated with the size of the deformable convolution kernel.

To alleviate the problems mentioned above caused by optical flow warping or single-size kernel compensation, we propose a novel Heterogeneous Deformable Compensation network for Video Compression (HDCVC). At first, the framework will calculate the accurate transformation relationship between the downsampled adjacent frame features. Thus to obtain the composition of different size deformable kernels, we design a scheme inspired by the idea\cite{singh2019hetconv} to generate learned heterogeneous deformable kernel offsets as motion information depending on the content in each spatial location. We design the Heterogeneous Deformable Convolution (HetDeformConv) to align the previous frame feature with the proposed kernel. Benefiting from the dynamic combination, HetDeformConv achieves stable performance and exhibits flexibility in handling complex motion, which will be discussed and analyzed in Section \ref{sec:ablation_HetDeform}. To transform motion information and residual into more compression-friendly representations and further eliminate statistical dependencies, we devise an effective scheme by remodeling the generalized divisive normalization (GDN)\cite{Balle17a} and propose the Spatial-Neighborhood-Conditioned Divisive Normalization (SNCDN) based on the representation extraction theory\cite{lyu2008nonlinear}. We also design a multi-frame enhanced reconstruction module to introduce context and temporal information assistance. The effectiveness of HDCVC is clearly shown as it outperforms the current state-of-the-art learned video compression approaches in terms of PSNR and MS-SSIM on most datasets. It also achieves competitive rate-distortion performance with the commercial codec x265 placebo preset\footnote{Placebo is the slowest setting in x265 which achieves the best quality among ten presets.}. In summary, the contributions of this paper can be summarized as follows:
\begin{figure*}[t]
	\centering
	\includegraphics[width=0.87\textwidth]{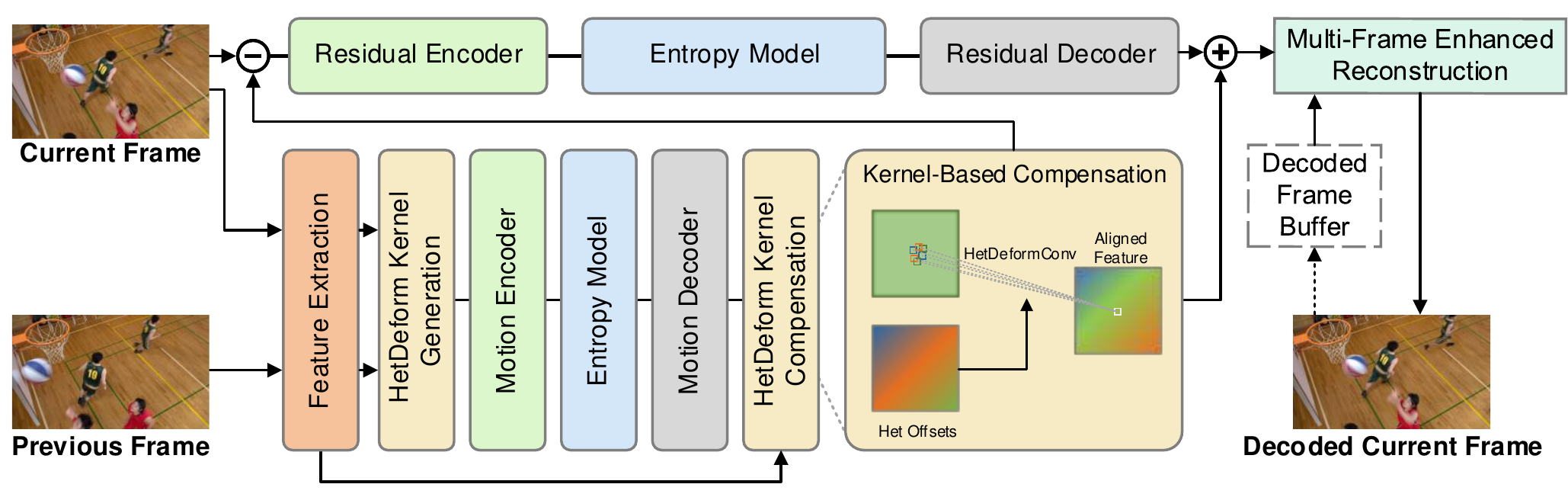}
	\caption{Overview of our proposed video compression framework HDCVC. When encoding, the framework extracts features from the previous frame $ \hat{I}_{t-1} $ and current frame $ I_{t} $, and generates heterogeneous deformable (HetDeform) kernel offsets $m_{t}$ for compensation. Both motion information and residual are compressed by the transformation autoencoder and quantized for entropy coding.}
	\label{fig:overview}
\end{figure*}
\begin{itemize}
	\item The proposed heterogeneous deformable compensation module can generate frame-content-adaptive kernel offsets and obtain dynamic composition of motion information. The proposed scheme is capable of achieving a robust rate-distortion performance in downsampled feature domain facilitated by the kernel diversity and the compensation's flexibility in handling various motion magnitudes.
	\item The novel Spatial Neighbor Conditioned Divisive Normalization is designed to achieve more efficient data transformation. Cooperated with the GDN, the SNCDN is capable of Gaussianizing data with less computation and improved compression performance.
	\item The multi-frame enhanced reconstruction module exploits valuable context and temporal information, which extracts target features from the current frame and context, aligns the previous frames in the feature domain, and fuses them for better frame reconstruction.
\end{itemize}

The remainder of this paper is organized as follows. Section \uppercase\expandafter{\romannumeral2} introduces the related work. In Section \uppercase\expandafter{\romannumeral3}, we present the overall framework with  detailed  discussions about HetDeform compensation, SNCDN and multi-frame enhanced reconstruction. In Section \uppercase\expandafter{\romannumeral4}, we show the performance of HDCVC and discuss the importance of each module, and then we conclude in Section \uppercase\expandafter{\romannumeral5}.
\section{Related Work}

\subsection{Learned Image Compression}
Recent years have witnessed performance improvement in the field of learned image compression. The pioneer work of Ball\'{e} \emph{et al.}\cite{Balle17a,balle2018variational} first proposed auto-encoder structures and optimized compression network in an end-to-end fashion. Besides, they also introduced generalized divisive normalization (GDN) to Gaussianize the input image into latent representation. After that, context-adaptive entropy models\cite{minnen2018joint,lee2018context} and Gaussian Mixture Likelihoods\cite{cheng2020learned} were designed for further improving the rate-distortion (RD) performance. Cheng \emph{et al.}\cite{cheng2019energy} provided a mathematical analysis on the energy compaction property and propose a new normalized coding gain metric to achieve high compression efficiency. Akbari\emph{et al.}\cite{9385968} propose a new variable-rate image compression framework, which adopts octave convolution and its variant. Ma \emph{et al.}\cite{8931632} propose a novel wavelet-like transform framework for natural image compression. Mei \emph{et al.}\cite{8931632} propose a end-to-end quality and spatial scalable image compression model, exploring the potential of feature-domain representation prediction and reuse. Most of the methods have outperformed the traditional image compression standards, such as JPEG\cite{wallace1992jpeg}, JPEG2000\cite{skodras2001jpeg}, BPG\cite{bellard2014bpg} and even intra mode of VVC\cite{bross2021developments}.

\subsection{Learned Video Compression}
In the past few years, learned video compression methods have attracted more attention among researchers. For instance, the LSTM-based approach\cite{Wu:2018} creatively treats the video compression task as frame interpolation and achieves comparable performance with H.264 (x264 LDP veryfast)\cite{x264}. However, the solution uses traditional block-based motion estimation and encodes the information by existing non-deep-learning-based coding methods. The Deep Video Compression (DVC)\cite{Lu2019CVPR} was proposed as the first end-to-end optimized video compression algorithm. Unlike the traditional codec framework, DVC replaced motion vectors with optical flow, making joint learning possible. Shortly afterward, Habibian \emph{et al.}\cite{habibian2019video} designed an autoencoder-based structure with auto-regressive prior. Both DVC and Habibian's approaches achieve comparable performance with H.265 (x265 LDP veryfast). Yang \emph{et al.}\cite{yang2020learning} designed the novel RNN-based compression method HLVC and utilized hierarchical quality layers for compressed video enhancement. Lu \emph{et al.}\cite{lu2020end} extended their DVC with various types of models to adapt to different application scenarios. Hu \emph{et al.}\cite{hu2020improving} adopted switchable multi-resolution representations for the flow maps based on the RD criterion. Liu \emph{et al.}\cite{9707786} proposed a hybrid motion compensation that considers both the pixel and feature-level alignment.

\subsection{Motion Estimation}
Considering computation complexity and bits consumption, block-based motion estimation and compensation are applied in hybrid coding framework\cite{h264,HEVC}. Though motion vectors are implementation-friendly and can effectively save bits, they are unsuitable for embedding in an end-to-end optimized compression system. Hence, DVC utilized a pre-trained flow prediction network\cite{ranjan2017optical} to avoid non-differentiable motion estimation. Liu \emph{et al.}\cite{liu2020learned} designed a one-stage unsupervised approach to estimate implicit flow and facilitate motion estimation. Agustsson\cite{agustsson2020scale} proposed a scale parameter for optical flow and achieved much more robust compensation performance. Recently, Hu \emph{et al.}\cite{hu2021fvc} utilized deformable convolution to accomplish compensation in the feature domain, alleviating the errors introduced by inaccurate pixel-level operations.

\section{Proposed Method}
\subsection{Notations and Overview}
Let $I_{t}$ and $\hat{I}_{t} (t \in {1, 2, ...}) $ denote a sequence of frames and its reconstructed version. From each frame $I_{t}$, the feature $ F_{t} $ is extracted for motion estimation and motion compensation. $ m_{t} $ is the motion information between $ \hat{F}_{t-1} $ and $ F_{t} $, and $ \hat{m}_{t}$ represents the reconstructed motion information. The compensation result $\tilde{I}^{com}_t$ is generated by the reference feature $\hat{F}_{t-1} $ and $ \hat{m}_{t} $. $ r_{t}$ is the residual between the predicted frame $\tilde{I}^{com}_t$ and the raw frame $I_{t} $. After obtaining the summation of $\tilde{I}^{com}_t$ and the reconstructed residual $ \hat{r}_{t}$, the final result $ \hat{I}_{t} $ is generated by multi-frame enhanced reconstruction with the aid of current result, context and previous frames. During encoding, $ m_{t} $ and $ r_{t} $ are transformed into latent representations ($ \tilde{m}_{t}$, $\tilde{r}_{t}$) and then compressed into the bitstream.

Fig.~\ref{fig:overview} shows an overview of the proposed network. The HDCVC focuses on encoding inter-frame motion information and residual into latent representation and high-quality reconstruction as a low-delay-prediction-oriented video compression model. The key procedure of our model is shown as follows:

\paragraph{Motion related module.}\label{sec:MotionEstimation}
Instead of embedding a complex estimation model, we design a simple fusion operation to estimate the heterogeneous deformable kernel offsets as motion information ${m}_{t}$. ${m}_{t}$ would be compressed into latent representation $\tilde{m}_{t}$ and reconstructed to $\hat{m}_{t}$ by the proposed motion transformation network.

The decoded motion information $\hat{m}_{t}$ is generated based on the frame content, with which the prediction frame $\tilde{I}^{com}_t$ is obtained by the proposed motion compensation operation HetDeformConv. Furthermore, a refinement network is cascaded after the compensation operation to restore resolution and dimension. We cover this in detail in Section \ref{sec:deformable-compensation}.

\paragraph{Residual related module.}
We compress residual $ r_{t} $ between the predicted frame $\tilde{I}^{com}_t$ and the raw frame $ I_{t} $ by the residual transformation network. The compressed representation $ \tilde{r}_{t} $ would be input into the bitstream along with the quantized motion representation. 
\begin{figure}[t]
	\centering
	\includegraphics[width=0.4\textwidth]{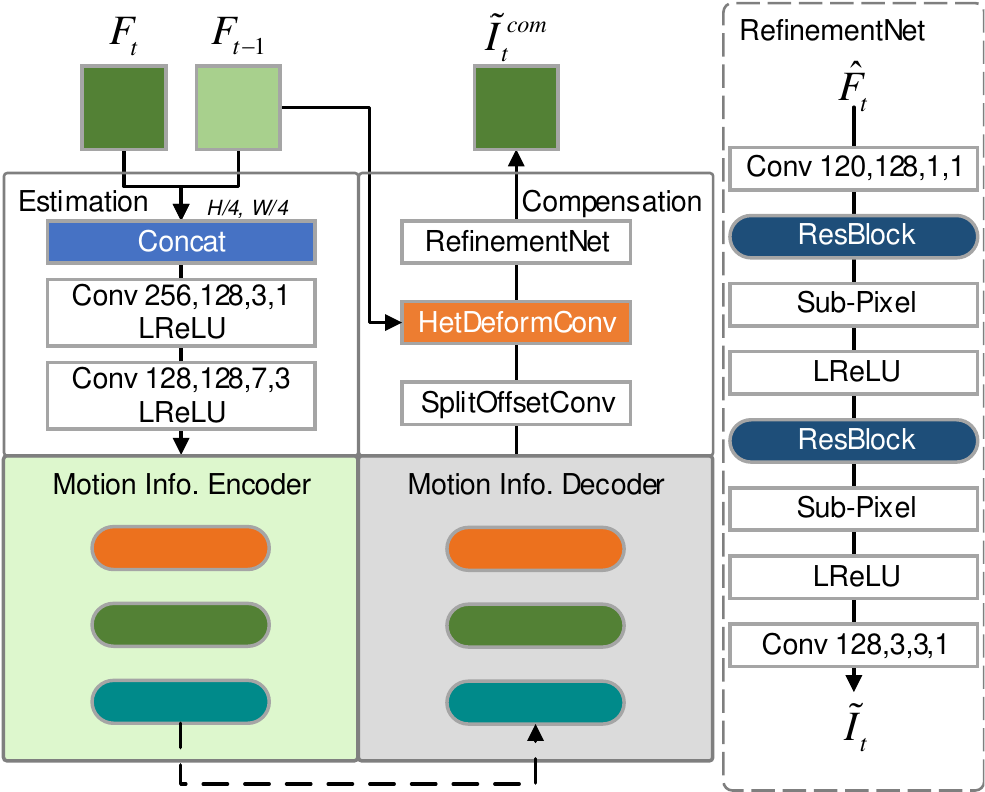}
	\caption{Overview of the motion estimation \& compensation network. We use Leaky ReLU as the activation function and denote it as LReLU. Conv M, N, K, S indicates a convolution layer with $K \times K$ size kernel, the number of input channels is M, the number of output channels is N, and the stride is S. The dashed line part in the figure omits the hyperprior module and the quantization operation.}
	\label{fig:motion_Compensation}
\end{figure}
\paragraph{Transform.}
We design an autoencoder-style transformation network for motion information and residual (Motion \& Residual Encoder/Decoder in Fig.~\ref{fig:overview}). These two networks have similar but different structures due to the discrepant feature sizes of motion and residual. We reduce the number of resampling operations for motion compression to guarantee the same output size of the two transformation networks. In order to accomplish latent representation transformation and obtain further compression improvement, we design a new divisive normalization scheme named SNCDN and combine residual blocks and the normalization series to form the non-linear transformer. More details regarding to the SNCDN are given in Section \ref{sec:encoder-decoder}. 

\paragraph{Entropy coding.}
We utilize hierarchical hyper prior and context-based prediction in our entropy model. In the training process, the module estimates probability distribution of each symbol and the bits cost of the output ($ \tilde{m}_{t} $, $ \tilde{r}_{t} $). The entropy module provides the adaptive context for accurate probability estimation during inference. The quantized representations are compressed into the bitstream by the arithmetic encoder.

\paragraph{Final reconstruction.}
For further quality enhancement, we generate the target feature with the decoded frame and extra context. Meanwhile, the reference frame features are aligned to the target feature with the proposed HetDeformConv. Finally, we fuse the target and aligned features with cascaded residual blocks and obtain the final reconstructed result. More details regarding the Multi-frame Enhanced Reconstruction are given in Section \ref{sec:MFER}. 

\subsection{Heterogeneous Deformable Compensation} \label{sec:deformable-compensation}
Fig.~\ref{fig:motion_Compensation} shows the detailed architecture of the motion estimation \& compensation network. Instead of embedding a pre-trained network or a complex estimation model, we only use two cascaded ordinary convolutions to generate motion information. After the compensation operation by Heterogeneous Deformable Convolution (HetDeformConv), a Refinement Network (RefinementNet) is designed for resolution restoration and alleviating the information loss caused by transformation and quantization. Given the reference frame $ \hat{I}_{t-1} $ and the current frame $ I_{t} $, the proposed approach will generate the combination of deformable kernel offsets of different sizes $ m_{t} $ between these two adjacent frames at the feature domain, and reconstruct the current frame using the HetDeformConv with the reference frame and decoded motion information $\hat{m}_{t}$.

\subsubsection{Feature extraction}
This module plays a role in extracting visual features from $\hat{I}_{t-1} $ and $ I_{t} $, which is composed of long-stride convolution layers and three cascaded residual blocks. During extraction, the input frames with the size of $H \times W \times 3$ are transformed and downsampled into corresponding features with the size of $H/4 \times W/4 \times 128$. The output of this module $F_{t-1}, F_{t}$ will be used for generating motion information.

\subsubsection{Motion estimation}
Without using a pre-trained optical flow estimation network like many other methods\cite{Lu2019CVPR,lu2020end,hu2020improving}, we apply a lightweight estimation network to transform the concatenated features into motion information $m_t$. We only use two ordinary convolutions to fuse the features of adjacent frames and generate the motion information $m_t$, which includes the Heterogeneous Deformable (HetDeform) kernel offsets. In our method, different from optical flow or single-size deformable kernel estimation, the combination of learnable offsets of various kernel sizes is estimated for each feature element position according to the frame content.
\begin{figure}[t]
	\centering
	\includegraphics[width=0.44\textwidth]{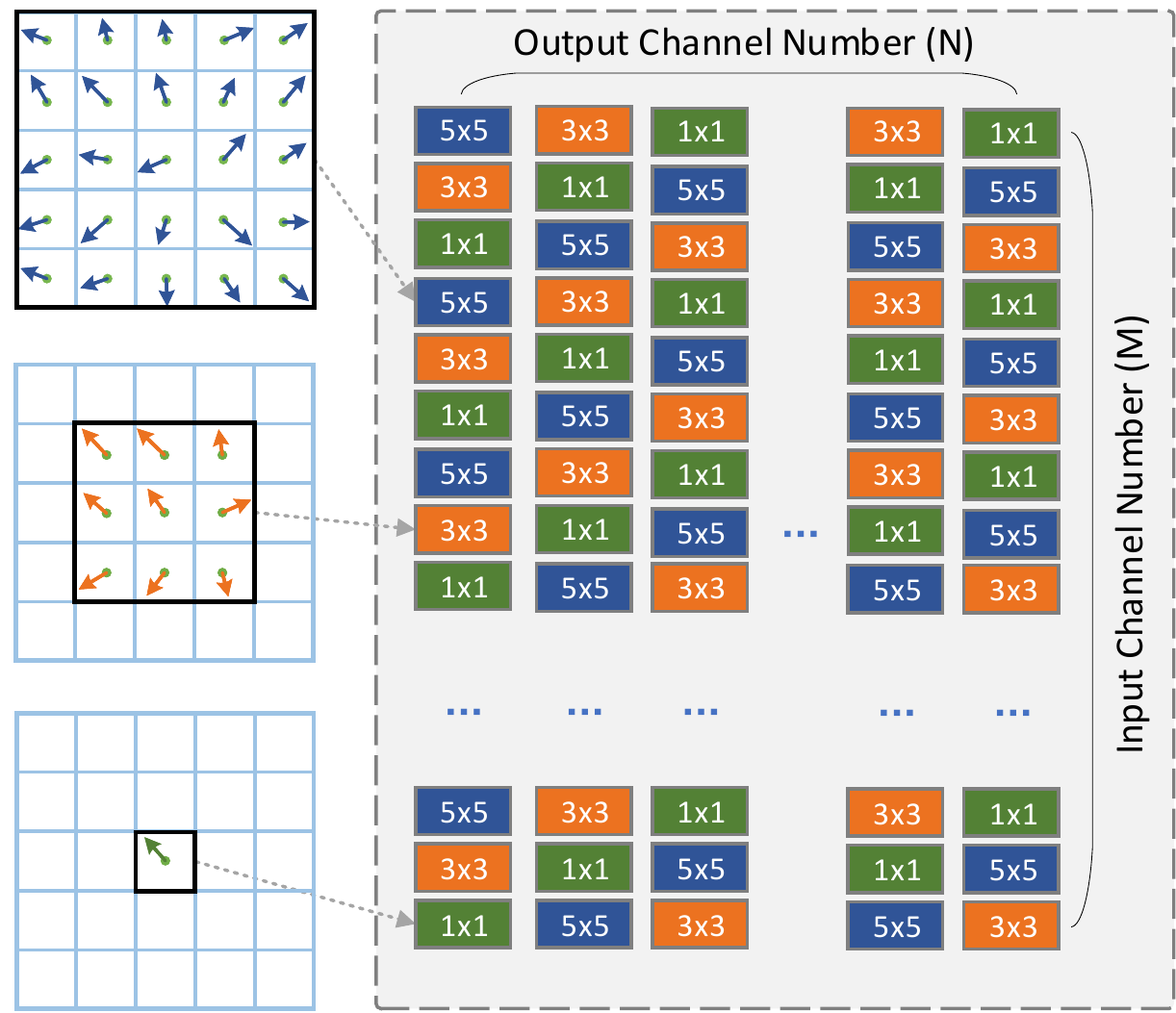}
	\caption{Heterogeneous Deformable Kernels. The learnable offsets will change adaptively with the spatial position. Still, to reduce the computational complexity and the network parameters, HDCVC shares offsets of the same kernel size in the grouped channel dimension.}
	\label{fig:het-kernel}
\end{figure}
\subsubsection{HetDeform kernel-based compensation}\label{sec:HetDeform Compensation}
Deformable convolution has demonstrated its advantage in feature extraction\cite{Dai:2017} and alignment \cite{tian2020tdan,wang2019edvr}. We expand the idea into the learned video compression framework and further resolve the non-globally optimal performance caused by different kernel sizes in downsampled feature domain. To reconstruct features of the current frames, the decoded motion information $ \hat{m}_{t}$ is obtained from the bitstream and can be separated into the dynamic Het offsets $\mathbf{p}_{k}^{Het}$. The formulation of single-size deformable convolution can refer to \cite{Dai:2017}. 

As for HetDeformConv, the dynamic combination of deformable kernels of three sizes (as Fig.~\ref{fig:het-kernel} shows) is utilized to predict the values of each feature point. We implement the HetDeformConv inspired by HetConv\cite{singh2019hetconv} to achieve operational efficiency and high-performance compensation. The core differences between the HetDeformConv and the ordinary deformable convolution is that ordinary version has only one size of deformable kernel, while HetDeformConv contains three different sizes of convolution kernels. Cooperated with the proposed kernel arrangement (See Fig.~\ref{fig:het-kernel}), the HetDeformConv has more diversity in kernels and offsets. The operation of HetDeformConv can be formulated in the following way:
\begin{equation}
	\begin{split}
		\hat{F}_{t}^{n}(\mathbf{p})=
		&\sum_{k_1=1}^{1\times1}\sum_{m_1=1}^{M/3}w_{k_1}\cdot {F}_{t-1}^{m_1}(\mathbf{p}+\mathbf{p}_{k_1}+\Delta\mathbf{p}_{k_1})\\
		&+\sum_{k_2=1}^{3\times3}\sum_{m_2=1}^{M/3}w_{k_2}\cdot {F}_{t-1}^{m_2}(\mathbf{p}+\mathbf{p}_{k_2}+\Delta\mathbf{p}_{k_2})\\
		&+\sum_{k_3=1}^{5\times5}\sum_{m_3=1}^{M/3}w_{k_3}\cdot {F}_{t-1}^{m_3}(\mathbf{p}+\mathbf{p}_{k_3}+\Delta\mathbf{p}_{k_3}),
	\end{split}
	\label{hetconv}
\end{equation} 
where $w_{k_i}, \mathbf{p}_{k_i}, \Delta\mathbf{p}_{k_i}, m_{i} (i\in1, 2, 3)$ are the kernel weights, the general sampling location, the additional learned offsets and the grouped input channel index for deformable kernel of different sizes, respectively. $t$ is the frame index of the sequence. $n (n\in1, 2, 3, ..., N)$ denotes the output channel index. $M$ and $N$ represent the input and the output channel number. As Equation \ref{hetconv} presents, we divide the input features into three parts by channel dimension and transform them separately with deformable convolution of different kernel sizes. 

For calculating computational cost, we denote $N\times H\times W$ as the size of the output feature map, and the total computational cost of single-size deformable convolution and HetDeformConv can be formulated by:
\begin{gather}
	ComCost_K= K \times K \times H\times W \times M \times N \times bilinear,
	\label{eq:single-size ComCost}\\
	ComCost_{Het}=\frac{35}{3} \times H\times W \times M \times N \times bilinear,
	\label{eq:HetKernel ComCost}
\end{gather}  

where $K$ is the size of ordinary deformable kernel and $35$ is computed by $1\times1 +3\times3+5\times5$. From Equation-\ref{eq:single-size ComCost}, \ref{eq:HetKernel ComCost} and Fig.~\ref{fig:comparison results}, there is evidence that the proposed method achieves global optimal result with the similar computation cost as $3\times 3$ deformable kernel compensation.
\begin{figure*}[t]
	\centering
	\includegraphics[width=0.82\textwidth]{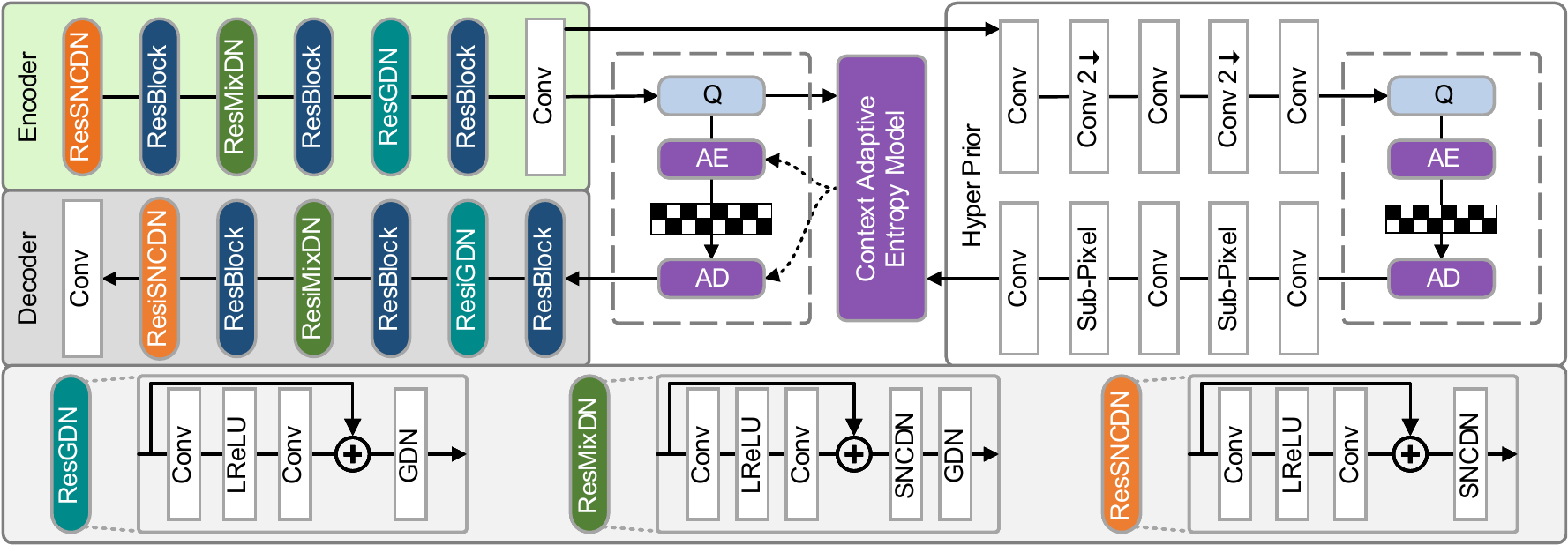}
	\caption{The architecture of our transformation network. The proposed ResDN series modules are embedded for transformation and inverse transformation on the encoder-decoder side. GDN indicates generalized divisive normalization\cite{balle2016density}. Q stands for quantization. IGDN, iMixDN and iSNCDN denote the inverse version of these divisive normalization methods.}
	\label{fig:motion_VAE}
\end{figure*}

\subsubsection{Refinement} 
Generally, motion compression procedure and quantization will cause non-negligible quality degradation. We apply an additional refinement module after the compensation operation to alleviate the information loss. Moreover, the compensation result will be restored to the original input size to facilitate residual calculation. Note that RefinementNet is designed mainly for resolution restoration, and its detailed architecture is shown in Fig.~\ref{fig:motion_Compensation}.

\subsection{Motion \& Residual Transformation}\label{sec:encoder-decoder}
The core task of the transformation network is to Gaussianize motion and residual into the compression-friendly latent representation, and previous work\cite{Balle17a,balle2018variational,minnen2018joint,lu2020end} has demonstrated the ability of generalized divisive normalization (GDN) in reducing data redundancy. Normalization can be modeled as the responses of neurons are divided by a common factor that typically includes the summed activity of a pool of neurons \cite{carandini2012normalization}, based on which the generalized form of divisive normalization\cite{balle2016density} is proposed.

\subsubsection{SNCDN} 
To reach the theoretical optimum of the entropy model performance, the distribution predicted by the model should be as close as possible to the actual marginal distribution of the latent representation. However, the statistic dependencies in the latent representation are hard to eliminate, thus leading to sub-optimal performance. Considering the strong correlation between the high-resolution feature point and its spatial neighborhood, we design the new divisive normalization scheme by extending the traditional work\cite{lyu2008nonlinear} to model the spatial neighborhood points as the neurons pool. The proposed Spatial-Neighborhood-Conditioned Divisive Normalization (SNCDN) will be applied for reducing statistical dependencies and effective Gaussianization when the features have not been continuously downsampled. GDN models the channel dimension of feature points as neighborhood, and our proposed SNCDN takes the spatial surrounding points of the target feature as divisive candidates. We denote the input $z$ of the $k$-th transform stage\footnote{We follow \cite{Balle17a} to denote the $k$-th transform stage as the $k$-th divisive normalization. In other words, the serial number is to distinguish the position of each divisive normalization operation.} at spatial location$(m,n)$ as $z^{(k)}(m,n)$, and the output as $y^{(k+1)}(m,n)$. The implementation of SNCDN can be formulated as:

\begin{equation}
	y^{(k+1)}(m,n)=\frac{z^{(k)}(m,n)}{(\beta^{(k)}+\sum_i \gamma^{(k)}|z_{i}^{(k)}(m,n)|^{\alpha})^{\epsilon}} ,
	\label{eq:SNCDN}	
\end{equation}
and the inverse version of SNCDN as inverse transform can be stated as:
\begin{equation}
	\hat{z}^{(k+1)}(m,n)={\hat{y}^{(k)}}(m,n) \cdot (\hat{\beta}^{(k)}+\sum_i \hat{\gamma}^{(k)}|\hat{y}_{i}^{(k)}(m,n)|^{\alpha})^{\epsilon},
	\label{eq:iSNCDN}
\end{equation}
where $\hat{y}^{(k)}, \hat{z}^{(k+1)}$ are the output and the input of the $k$-th inverse transform stage respectively. $z_{i}^{(k)}(m,n)$ denotes the neighboring pixels of spatial location$(m,n)$, $i$ stands for the spatial neighborhood index, $\phi= \left\{ \beta, \gamma \right\}, \theta= \left\{ \hat{\beta}, \hat{\gamma} \right\}$ are learnable parameters and $\alpha, \epsilon$ are set to 2 and $\frac{1}{2}$ by default.

\subsubsection{ResDN series} 
As shown in Fig.~\ref{fig:motion_VAE}, different from several previous work where GDN is directly embedded in the network, we unite the residual block\cite{he2016deep}, SNCDN and GDN as basic transform blocks and apply the transform block series in a cascaded style. As discussed above, SNCDN should implement before features are downsampled multiple times, so the location of ResSNCDN, ResMixDN and ResGDN are deployed according to this guideline. We have done ample experiments to prove the reliability of this arrangement. The motion and residual transformation networks are designed as similar but not precisely the same architectures due to the mismatched size between the input data. After the proposed transformation, the quantized components will be compressed into the bitstream by arithmetic coding. 

\subsection{Multi-Frame Enhanced Reconstruction}\label{sec:MFER}
Inspired by the recent work\cite{hu2021fvc} using the multi-reference-frame feature for final fusion, we design a Multi-Frame Enhanced Reconstruction (MFER) to further exploit context and temporal information. The detailed network architecture is shown in Fig. \ref{fig:multi-frame_enhancement}. We first generate reference features from the previous three frames $\hat{I}_{t-3}$, $\hat{I}_{t-2}$ and $\hat{I}_{t-1}$. After that, we treat the decoded frame, the predicted result and other related content as context\cite{li2021deep}, and fuse the context as target feature. The feature extraction network are composed of two cascaded convolution layers for downsampling and extracting feature. After generating the target feature and the reference features, we estimate the motion information between the target and the three reference features, and align the reference features to the target feature using the proposed HetDeform Convolution described in Section \ref{sec:HetDeform Compensation}, for exploiting additional temporal information. Finally, the cascaded resblocks and sub-pixel operations are applied for final enhancement and reconstruction. The MFER plays a similar role as the in-loop filter in the traditional codec, and it can alleviate error propagation and boost reconstruction quality without needing extra bits.

\subsection{Loss Function}\label{sec:loss}
We utilize a progressive training scheme to stabilize the optimization. Specifically, our framework focuses on stabilizing compensation performance in the early stages of model optimization, and thus only motion related modules are engaged in training. We apply the following rate-compensation loss $L_{r-c}$ in this stage:
\begin{figure}[t]
	\centering
	\includegraphics[width=0.42\textwidth]{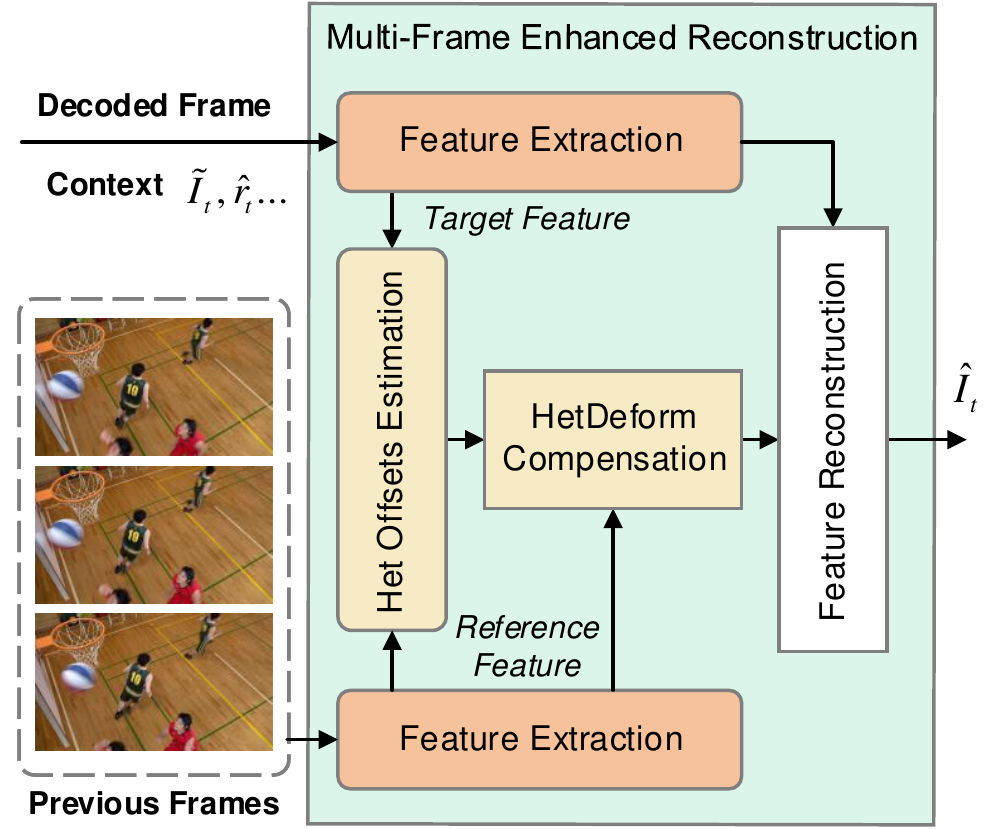}
	\caption{The architecture of the proposed multi-frame enhanced reconstruction module.}
	\label{fig:multi-frame_enhancement}
\end{figure}
\begin{equation}
	L_{r-c}=\lambda \cdot D_{com}+R=\lambda \cdot d(\tilde{I}^{com}_t,I_{t})+R_{\tilde{m}},
	\label{equa:MCC_loss}
\end{equation}
in which $\tilde{I}^{com}_t$ denotes the motion compensation results, and $\lambda$ is the Lagrange multiplier that determines the trade-off between the entropy estimation results $ R $ of latents and hyper-latents and the compensation distortion $D_{com}$. Note that in this training period, $D_{com}$ only denotes mean-square error (MSE) in our PSNR and MS-SSIM\cite{MS-SSIM} models.

In the second stage, we optimize motion related modules and residual related modules together. The goal of our video compression framework is to minimize the number of bits used for encoding the video, while reducing distortion between the original input frame $I$ and the reconstructed frame $\hat{I}$. Therefore, we propose the total RD optimization loss function:
\begin{equation}
	L_{total}=\lambda \cdot D+R=\lambda \cdot d(\hat{I}_{t},I_{t})+R_{\tilde{m}}+R_{\tilde{r}}
	\label{equa:total_loss}
\end{equation}
where $d(I_{t},\hat{I}_{t}) $ denotes the distortion between $I_{t} $ and $\hat{I}_{t} $. In the final stage, we train the whole framework including Multi-Frame Enhanced Reconstruction jointly. To train our PSNR model, we use MSE as the distortion function, and as to the MS-SSIM (Multi-scale Structural Similarity) model, we use $d(x, y)=1-$ MS-SSIM$(x,y)$ as the distortion function.

\begin{figure*}[t!]
	\centering
	\begin{subfigure}{.32\textwidth}
		\centering
		\includegraphics[width=\textwidth]{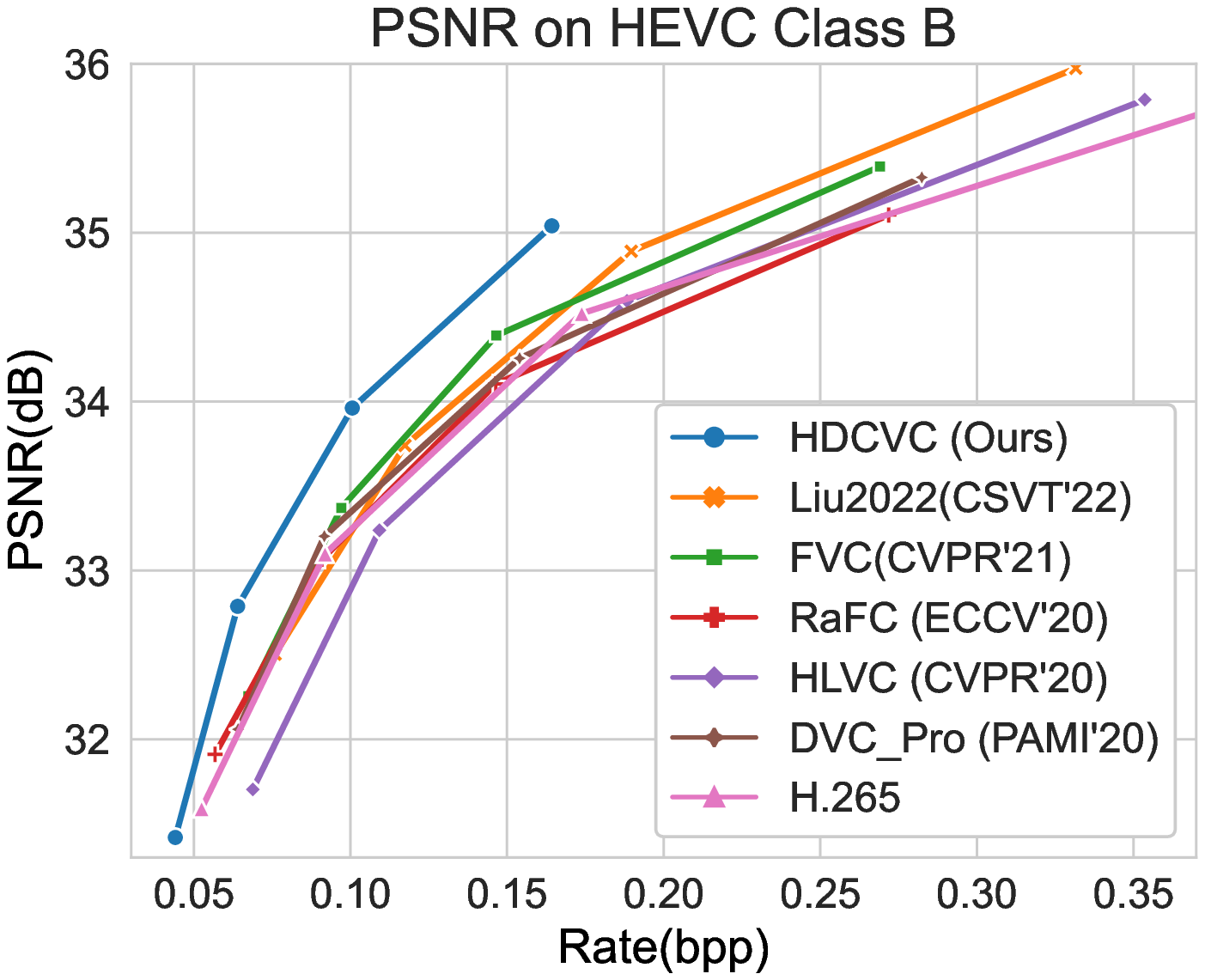}
		\label{ClassB_PSNR}
	\end{subfigure}
	\begin{subfigure}{.32\textwidth}
		\centering
		\includegraphics[width=\textwidth]{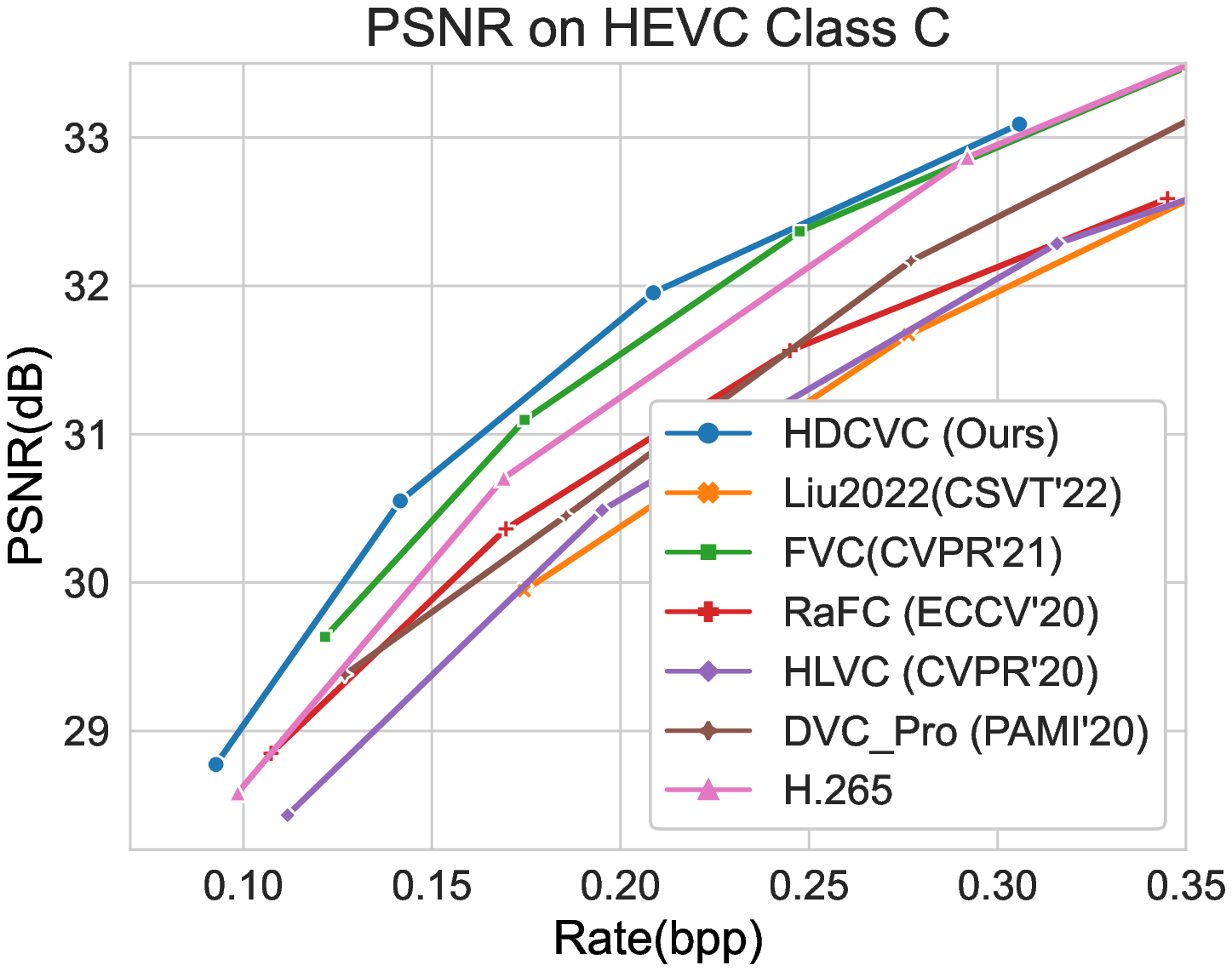}
		\label{ClassC_PSNR}
	\end{subfigure}
	\begin{subfigure}{.32\textwidth}
		\centering
		\includegraphics[width=\textwidth]{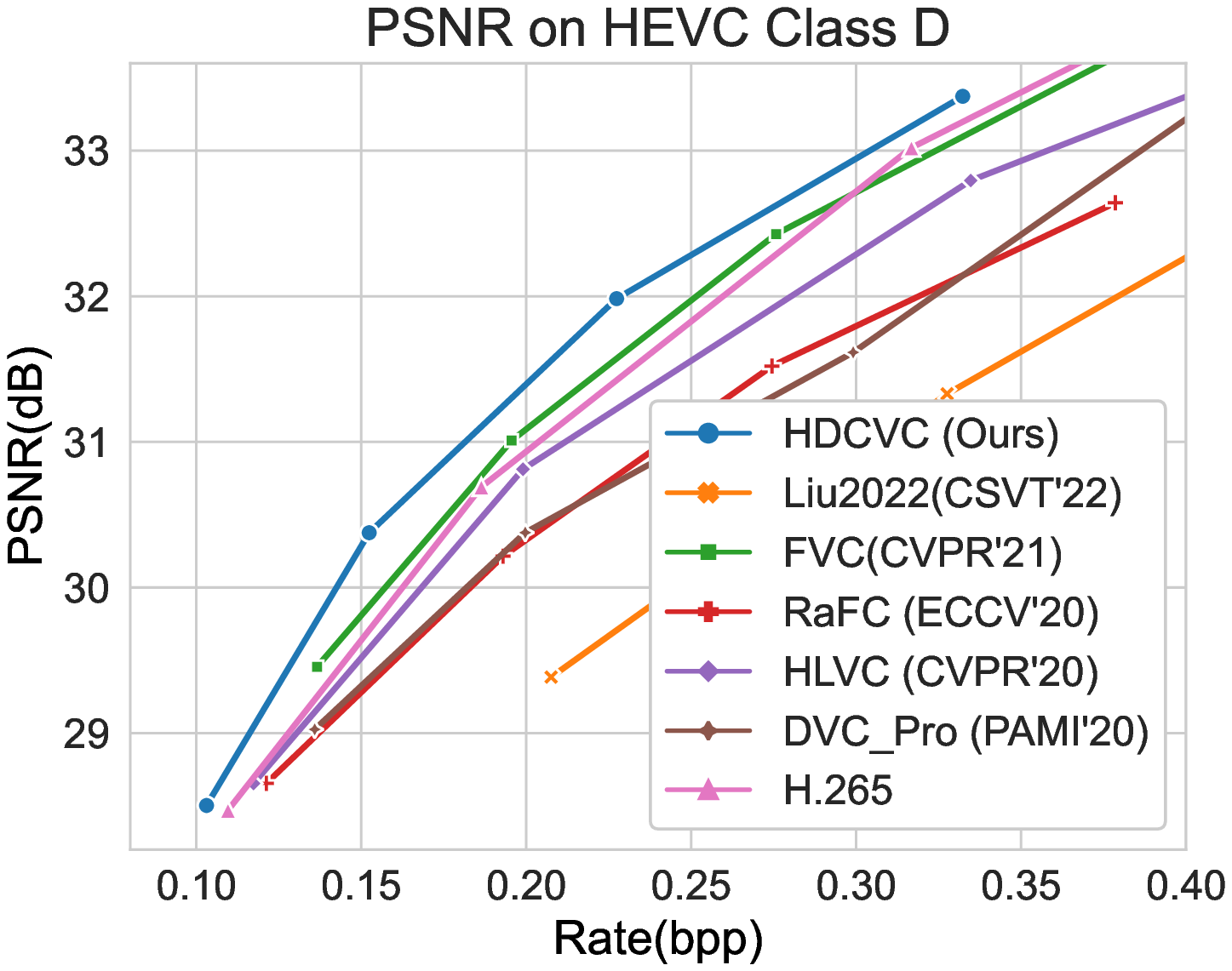}
		\label{ClassD_PSNR}
	\end{subfigure}
	\\
	\begin{subfigure}{.32\textwidth}
		\centering
		\includegraphics[width=\textwidth]{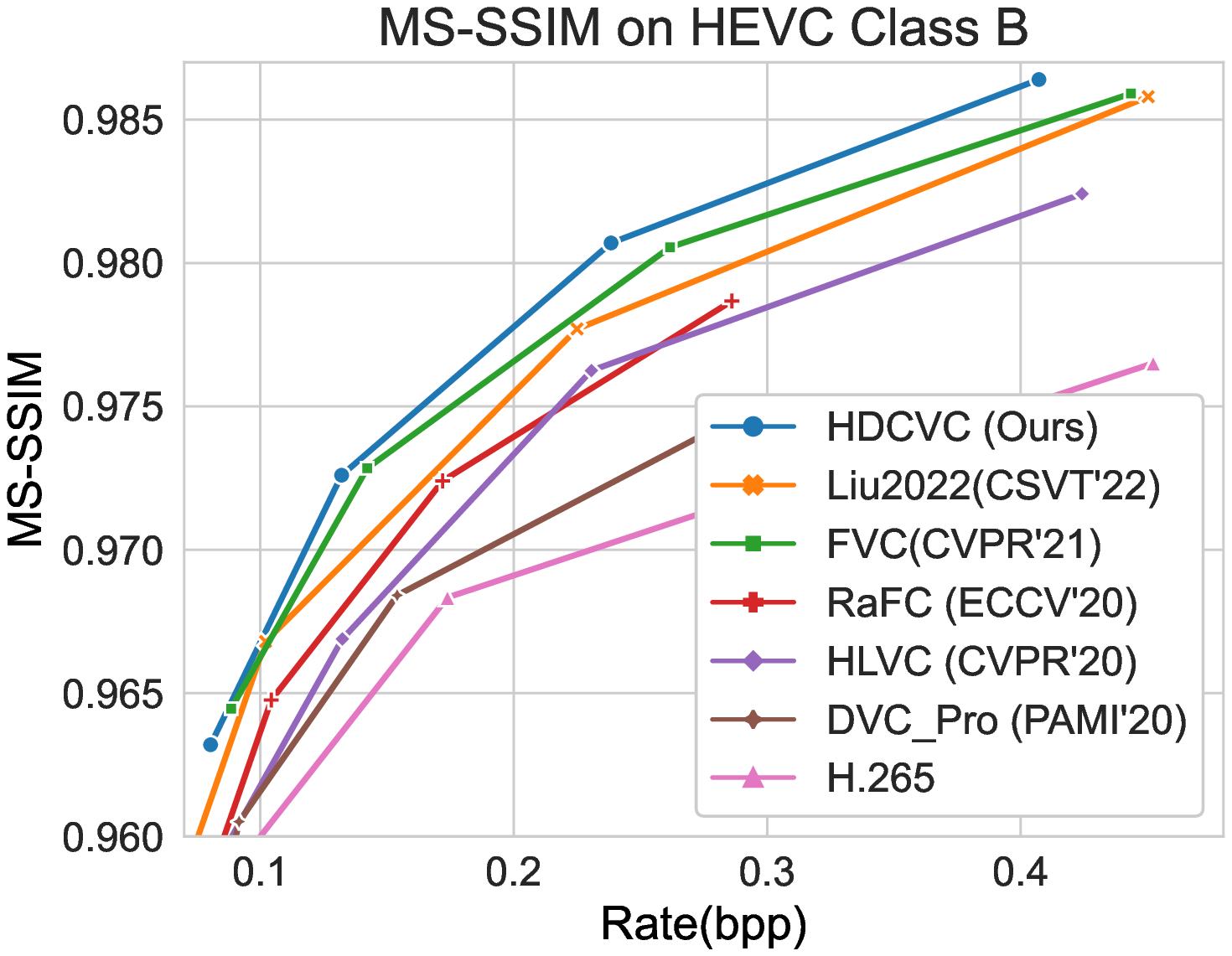}
		\label{ClassB_MSSSIM}
	\end{subfigure}
	\begin{subfigure}{.32\textwidth}
		\centering
		\includegraphics[width=\textwidth]{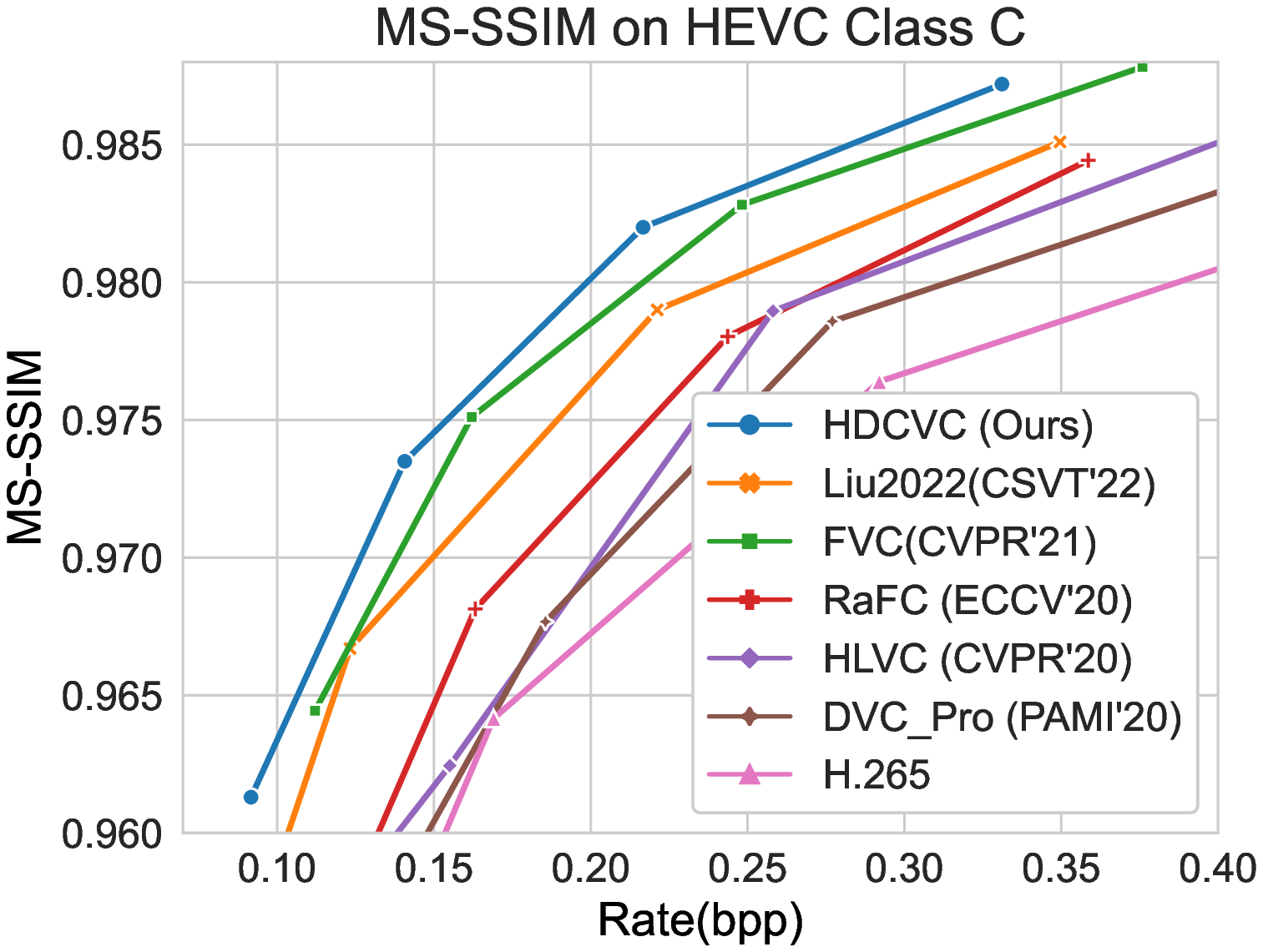}
		\label{ClassC_MSSSIM}
	\end{subfigure}
	\begin{subfigure}{.32\textwidth}
		\centering
		\includegraphics[width=\textwidth]{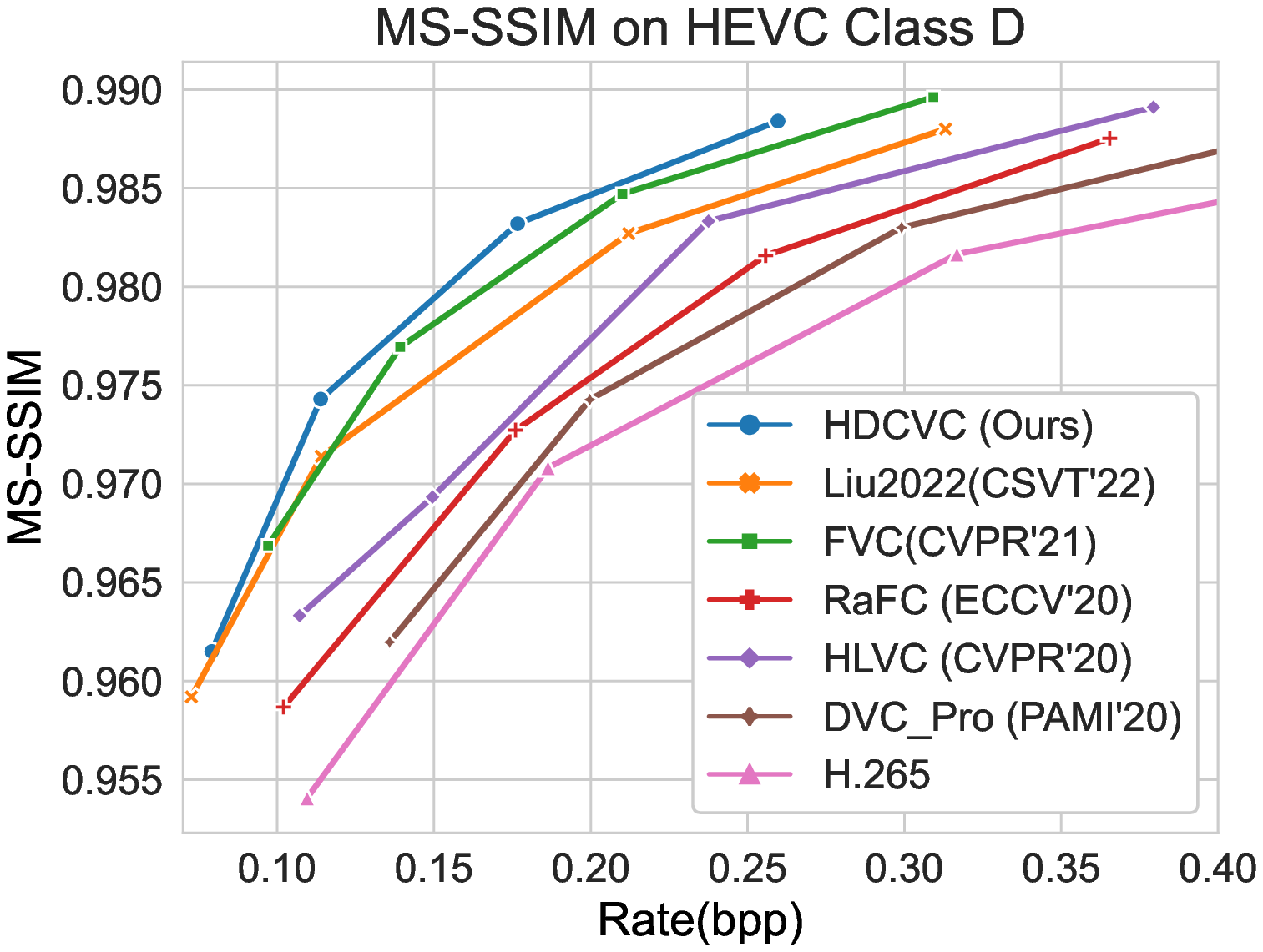}
		\label{ClassD_MSSSIM}
	\end{subfigure}
	\\
	\begin{subfigure}{.32\textwidth}
		\centering
		\includegraphics[width=\textwidth]{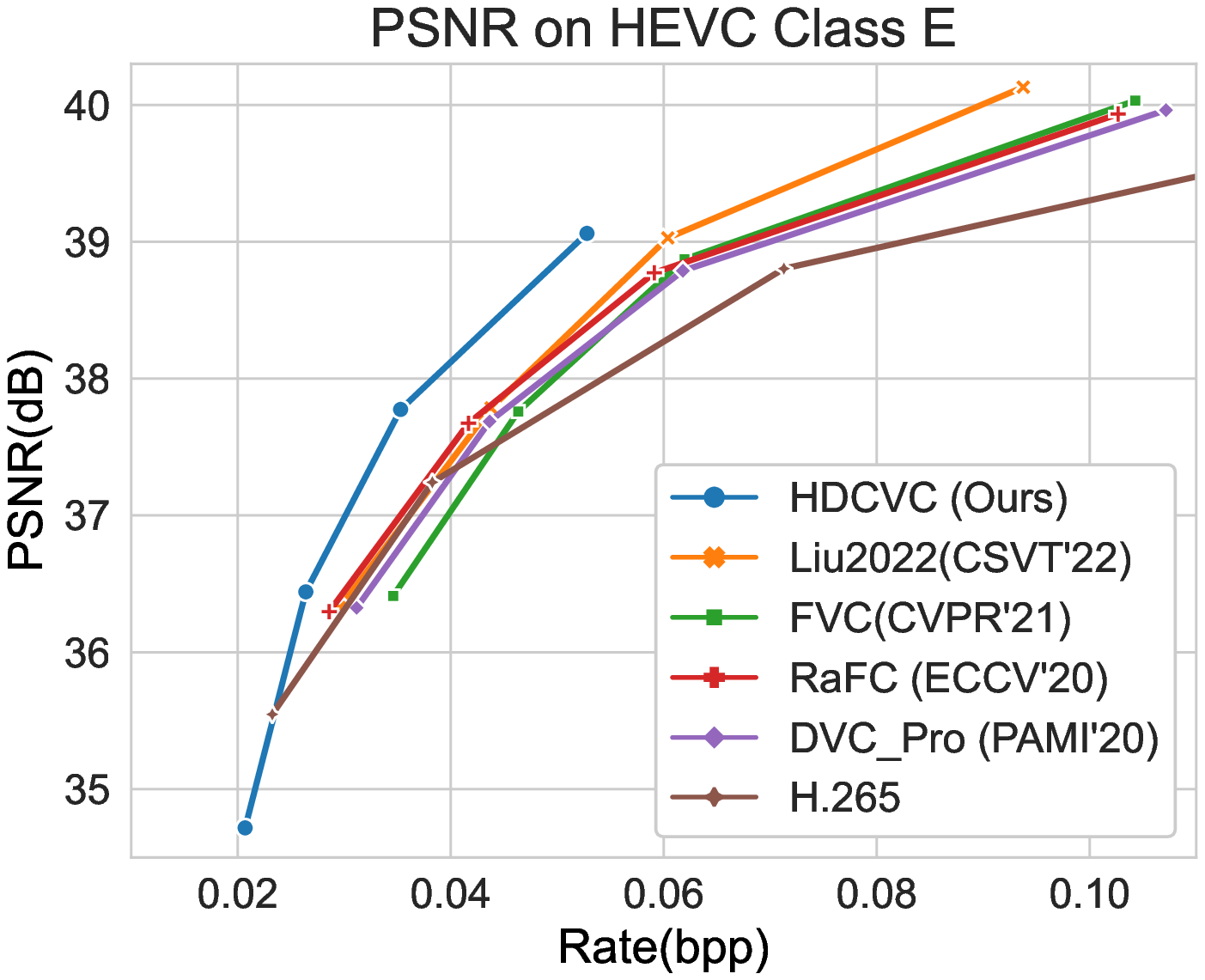}
		\label{ClassE_PSNR}
	\end{subfigure}
	\begin{subfigure}{.32\textwidth}
		\centering
		\includegraphics[width=\textwidth]{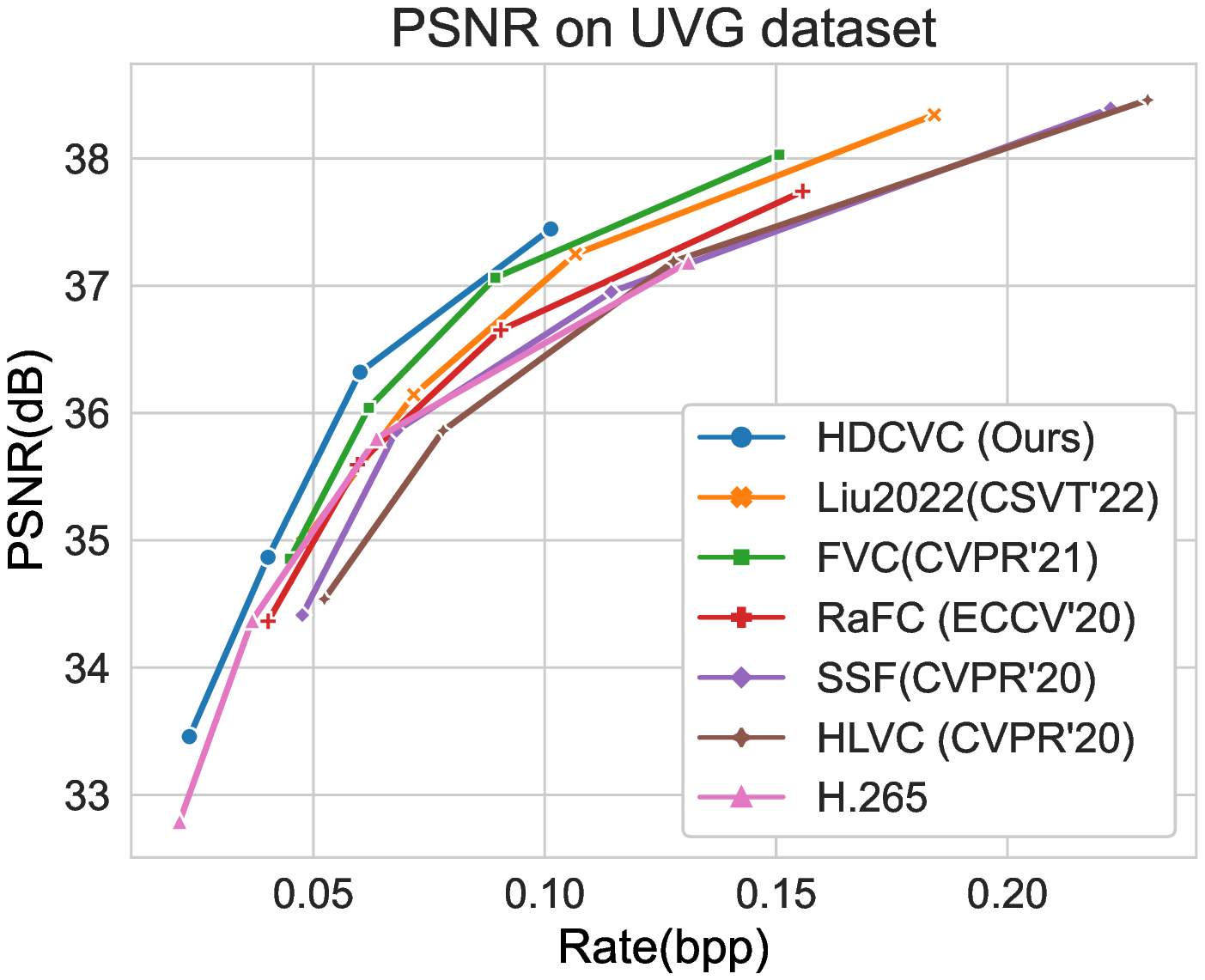}
		\label{UVG_PSNR}
	\end{subfigure}
	\begin{subfigure}{.32\textwidth}
		\centering
		\includegraphics[width=\textwidth]{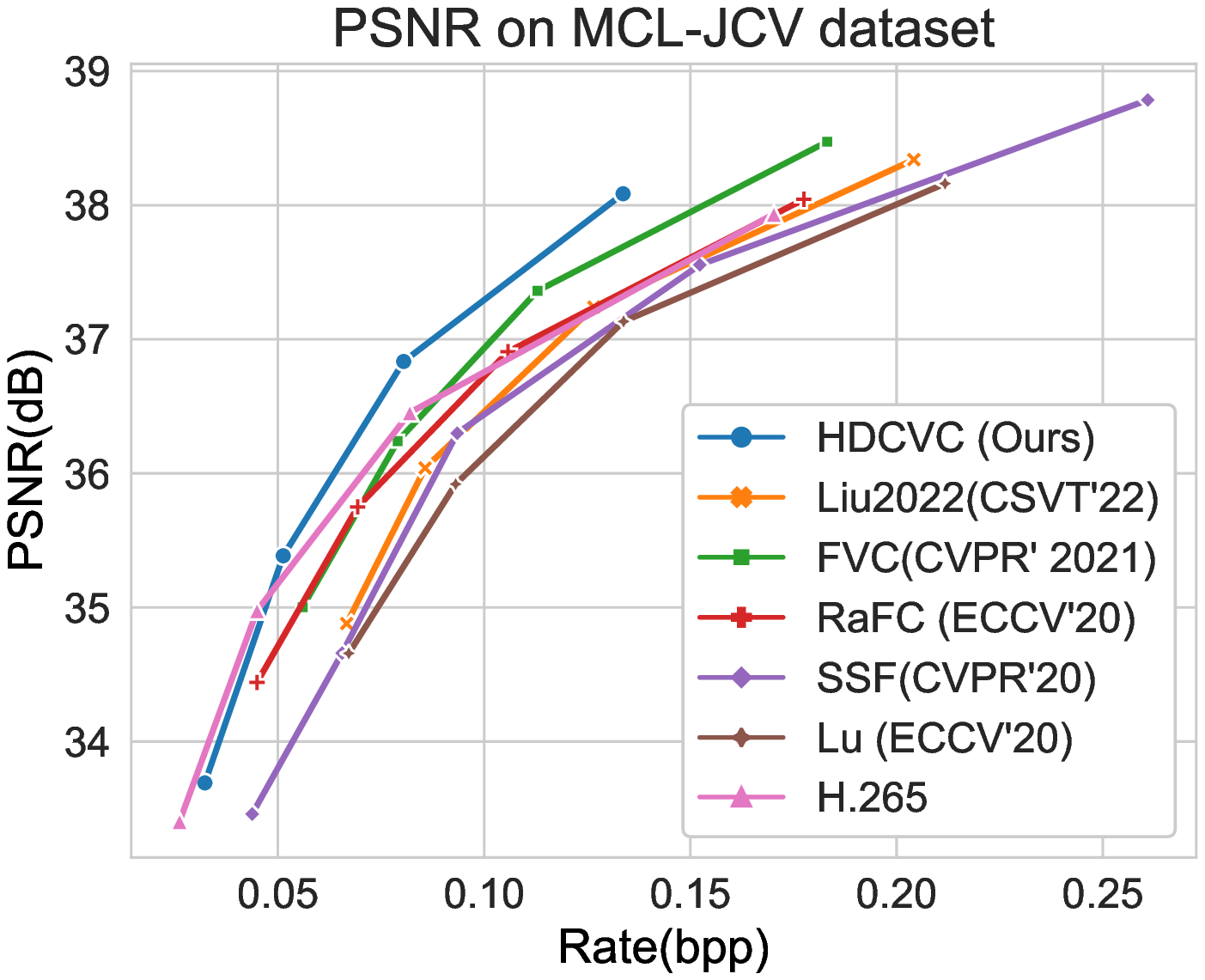}
		\label{MCLJCV_PSNR}
	\end{subfigure}
	\\
	\begin{subfigure}{.32\textwidth}
		\centering
		\includegraphics[width=\textwidth]{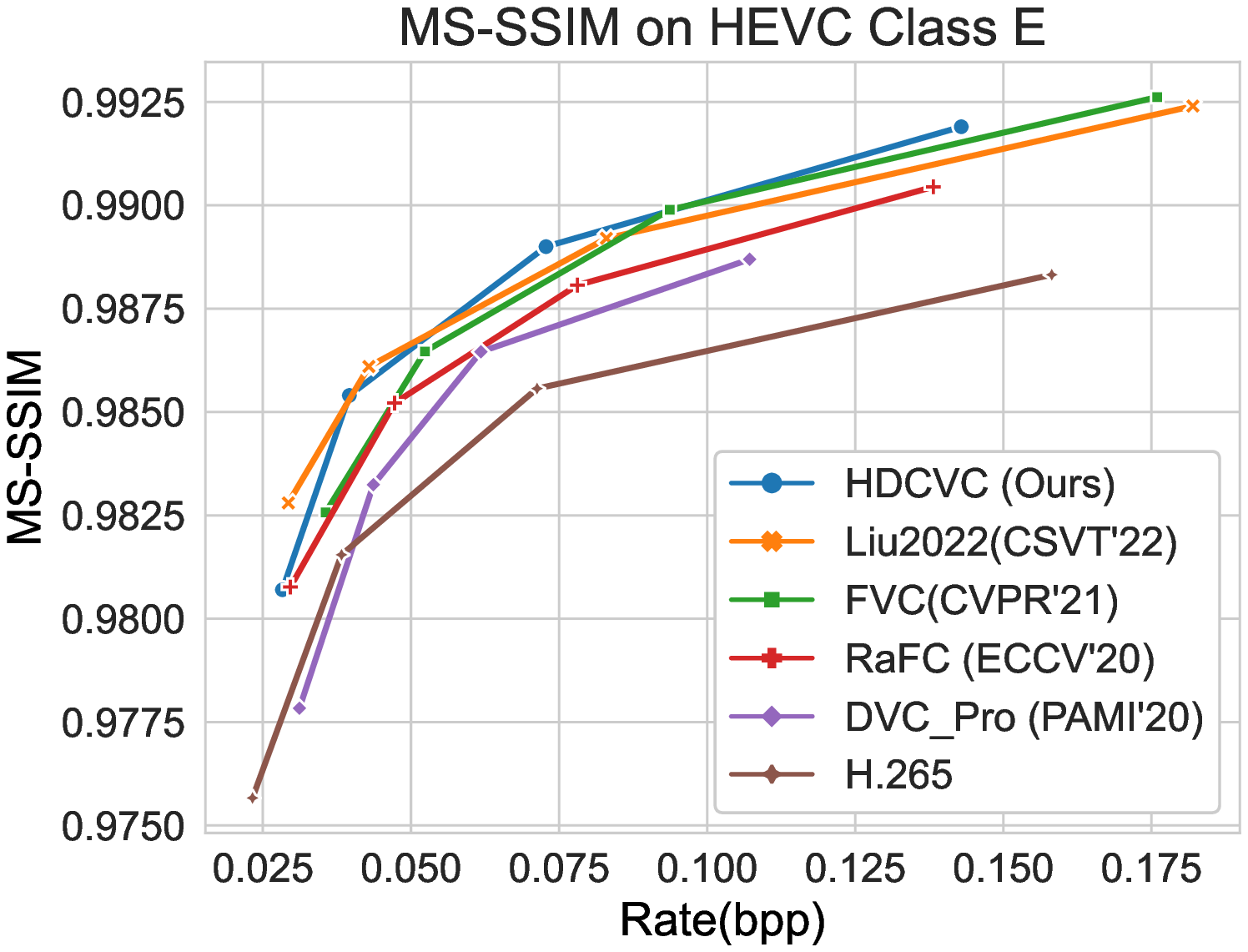}
		\label{ClassE_MSSSIM}
	\end{subfigure}
	\begin{subfigure}{.32\textwidth}
		\centering
		\includegraphics[width=\textwidth]{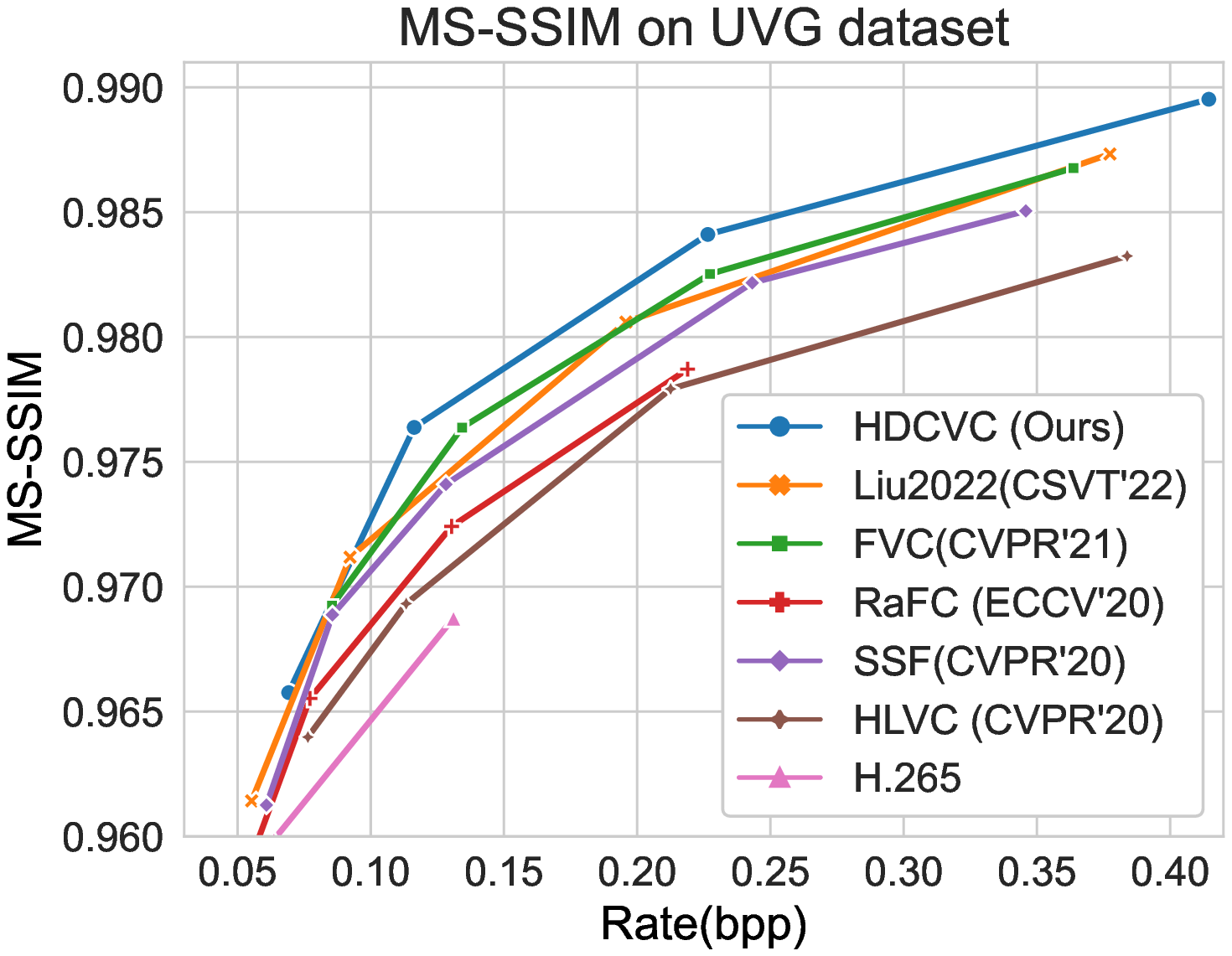}
		\label{UVG_MSSSIM}
	\end{subfigure}
	\begin{subfigure}{.32\textwidth}
		\centering
		\includegraphics[width=\textwidth]{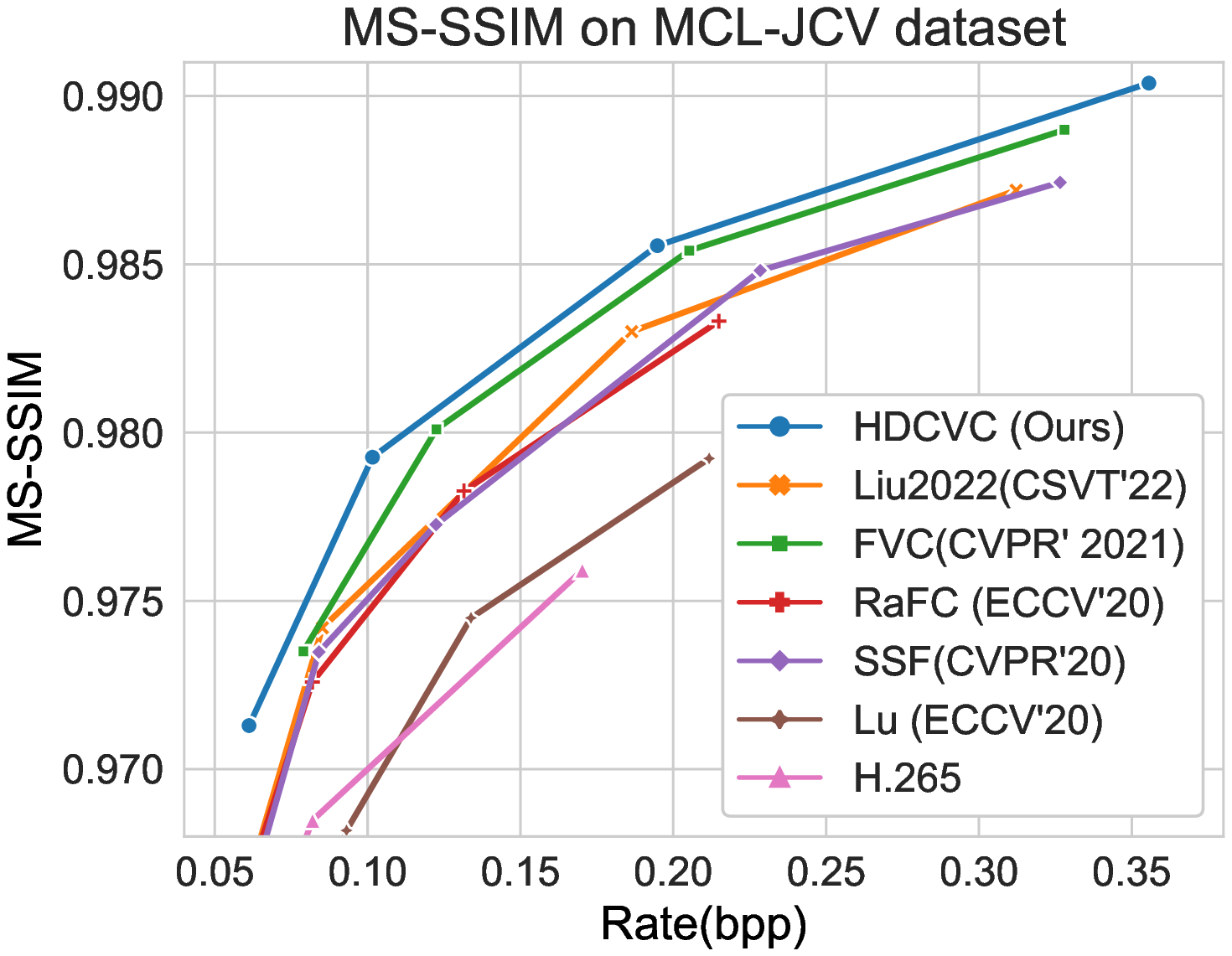}
		\label{MCLJCV_MSSSIM}
	\end{subfigure}
	\caption{The rate-distortion curve of our approach compared with H.265 (x265 LDP placebo) and the recent learned video compression approaches on the HEVC, UVG and MCL-JCV sequences.}
	\label{Result}
\end{figure*}
\section{Experiments}
\subsection{Datasets and Implementation Details}
\subsubsection{Datasets}
To train our model, we adopted the dataset provided by \cite{xue:2019}, which is collected from the Vimeo and released named Vimeo-90K. The dataset contains 89,800 video clips with various and complex real-world motions. Each sample contains seven consecutive frames with a fixed resolution of $448 \times 256$. To validate the effectiveness of our model and compare compression performance with other approaches, we test our models on four datasets: the HEVC dataset\cite{bossen2013common}, the UVG dataset\cite{UVG} and the MCL-JCV dataset \cite{wang2016mcl}. The resolution of these datasets is diverse from $352\times288$ to $1920\times1080$, which can comprehensively evaluate the performance of compression models.

\subsubsection{Training settings}
As HDCVC requires reconstructed frame $\hat{I}_{t-1} $ for motion compensation during training, we train our model six times in succession using seven consecutive frames by buffering the previous reconstructed frame as \cite{Lu2019CVPR}. With regard to I-frame compression, we choose cheng2020-anchor\cite{cheng2020learned} from CompressAI\cite{begaint2020compressai} to achieve higher rate-distortion performance. 

We train eight models with different $\lambda $ to control RD performance ($\lambda $ = 256, 512, 1024, 2048 for the PSNR model and $\lambda $ = 8, 16, 32, 64 for the MS-SSIM model). The batch size is set to 16, and we randomly crop the training sequences into the resolution of $256 \times 256$. The Adam optimizer is adopted whose parameters $\beta_{1}$ and $\beta_{2}$ are set as $0.9$ and $0.999$. The learning rate is initialized to $1 \times 10^{-4}$ and decreased by a factor of 2 when evaluation performance becomes stable. The entire network converges after 550000 iterations. All experiments are conducted using the PyTorch with NVIDIA RTX 3090 GPUs.

\subsection{Experimental setting}
Our model's intra period is set to 10 for the HEVC dataset\cite{bossen2013common} and 12 for other datasets (the UVG\cite{UVG} and MCL-JCV\cite{wang2016mcl}) for a fair comparison with other learned video compression methods. Besides, our framework requires the input frame's width and height to be a multiple of 64, so we crop the frame margin to meet the model requirements as previous work does.

In order to compare rate-distortion performance with hybrid video coding standards\cite{h264,HEVC}, we refer to the setting in \cite{Lu2019CVPR,yang2020learning} to generate the bitstream of x264 and x265 with FFmpeg\cite{x264,x265}. Please refer to \cite{yang2020learning} for the detailed configurations of x264, x265.
\begin{table*}[t]
	\begin{center}
		\caption{BDBR (\%) Results with the anchor of H.265 (x265 LDP placebo). \textbf{\textcolor{red}{Red}} and \textcolor{blue}{Blue} indicates the best and the second-best performance, respectively. All RD results of DVC, DVC\_Pro, HLVC, Hu, Liu2022 and FVC are provided by their authors.}
		\label{BDBR}
		\scalebox{0.88}{
			\begin{tabular}{c||ccccccc|ccccccc}
				\toprule[1.2pt]
				& \multicolumn{7}{c|}{BDBR (\%) calculated by PSNR}                                                                                                                           & \multicolumn{7}{c}{BDBR (\%) calculated by MS-SSIM}                                                                                                                         \\ \cmidrule{2-15}
				& \begin{tabular}[c]{@{}c@{}}DVC\\ \cite{Lu2019CVPR}\end{tabular}     & \begin{tabular}[c]{@{}c@{}}DVC\_Pro\\ \cite{lu2020end}\end{tabular} & \begin{tabular}[c]{@{}c@{}}HLVC\\ \cite{yang2020learning}\end{tabular}          & \begin{tabular}[c]{@{}c@{}}Hu\\ \cite{hu2020improving}\end{tabular}           & \begin{tabular}[c]{@{}c@{}}Liu2022\\ \cite{9707786}\end{tabular}                     & \begin{tabular}[c]{@{}c@{}}FVC\\ \cite{hu2021fvc}\end{tabular}                   & \begin{tabular}[c]{@{}c@{}}HDCVC\\ (Ours)\end{tabular} & \begin{tabular}[c]{@{}c@{}}DVC\\ \cite{Lu2019CVPR}\end{tabular}     & 
				\multicolumn{1}{c|}{\begin{tabular}[c]{@{}c@{}}DVC\_Pro\\ \cite{lu2020end}\end{tabular}} & \begin{tabular}[c]{@{}c@{}}HLVC\\ \cite{yang2020learning}\end{tabular}     &  \begin{tabular}[c]{@{}c@{}}Hu\\ \cite{hu2020improving}\end{tabular}       &\begin{tabular}[c]{@{}c@{}}Liu2022\\ \cite{9707786}\end{tabular}       & \begin{tabular}[c]{@{}c@{}}FVC\\ \cite{hu2021fvc}\end{tabular}                                      & \begin{tabular}[c]{@{}c@{}}HDCVC\\ (Ours)\end{tabular} \\ \cmidrule{2-15}
				\multirow{-5}{*}{Dataset}                                 & \multicolumn{9}{c|}{Optimized for PSNR}    & \multicolumn{5}{c}{Optimized for MS-SSIM}                                                                                         \\ 
				\midrule[1.2pt]
				HEVC Class B    & 25.55  & -0.24   & 10.09    & 2.39               & \textcolor{blue}{-6.95}
				          &  -5.99          & \textcolor{red}{ \textbf{-22.08 }}               & 17.81  & \multicolumn{1}{c|}{-7.13 }  & -22.36      & -24.86& -35.72 & \textcolor{blue}{ -44.32 }          & \textcolor{red}{ \textbf{-45.96 }}               \\ \midrule
				HEVC Class C  & 40.08  & 13.16   & 24.01     & 11.93            & 27.27
				              & \textcolor{blue}{ -4.54 } & \textcolor{red}{ \textbf{-10.59} }                          & 4.45   & \multicolumn{1}{c|}{-7.93 }  & -13.45   & -18.86 & -33.75  & \textcolor{blue}{ -40.15 }          & \textcolor{red}{ \textbf{-44.25 }}               \\ \midrule
				HEVC Class D   & 39.70  & 20.86   & 8.18     & 17.34    & 47.44
				                      & \textcolor{blue}{-1.10 } & \textcolor{red}{ \textbf{ -9.49 }}                          & 4.56   & \multicolumn{1}{c|}{-8.44 }  & -24.05   & -16.01  & -39.58 & \textcolor{blue}{ -43.54 } & \textcolor{red}{ \textbf{-47.00} }                        \\ \midrule
				HEVC Class E    & 14.72  & -6.54   & -         &  -10.47 & \textcolor{blue}{-14.32}
				  & -6.21                                  & \textcolor{red}{ \textbf{-19.51 }}               & 6.62   & \multicolumn{1}{c|}{-11.38 } & -            & -28.92 & \textcolor{red}{\textbf{-45.58}}				   &  -35.99           & \textcolor{blue}{-39.82}               \\ \midrule
				UVG      & 26.73  & -        & 17.30    & -0.15          & -5.46
				                & \textcolor{blue}{ -13.28 }         & \textcolor{red}{ \textbf{-17.30 }}               & 32.30  & \multicolumn{1}{c|}{-}        & -20.55    & -22.57& -31.14  & \textcolor{blue}{ -44.39 }          & \textcolor{red}{ \textbf{-48.09 }}               \\ \midrule
				MCL-JCV  & -       & -        & -         & 10.07          & 20.08  & \textcolor{blue}{ 1.84 }           & \textcolor{red}{ \textbf{-7.83 }}                & -       & \multicolumn{1}{c|}{-}        & -           & -25.94& -30.23& \textcolor{blue}{ -43.45 }          & \textcolor{red}{ \textbf{-48.16 }}               \\ \bottomrule[1.2pt]
			\end{tabular}
		}
	\end{center}
	
\end{table*}
\begin{figure*}[t]
	\centering
	\begin{subfigure}{.32\textwidth}
		\centering
		\includegraphics[width=\textwidth]{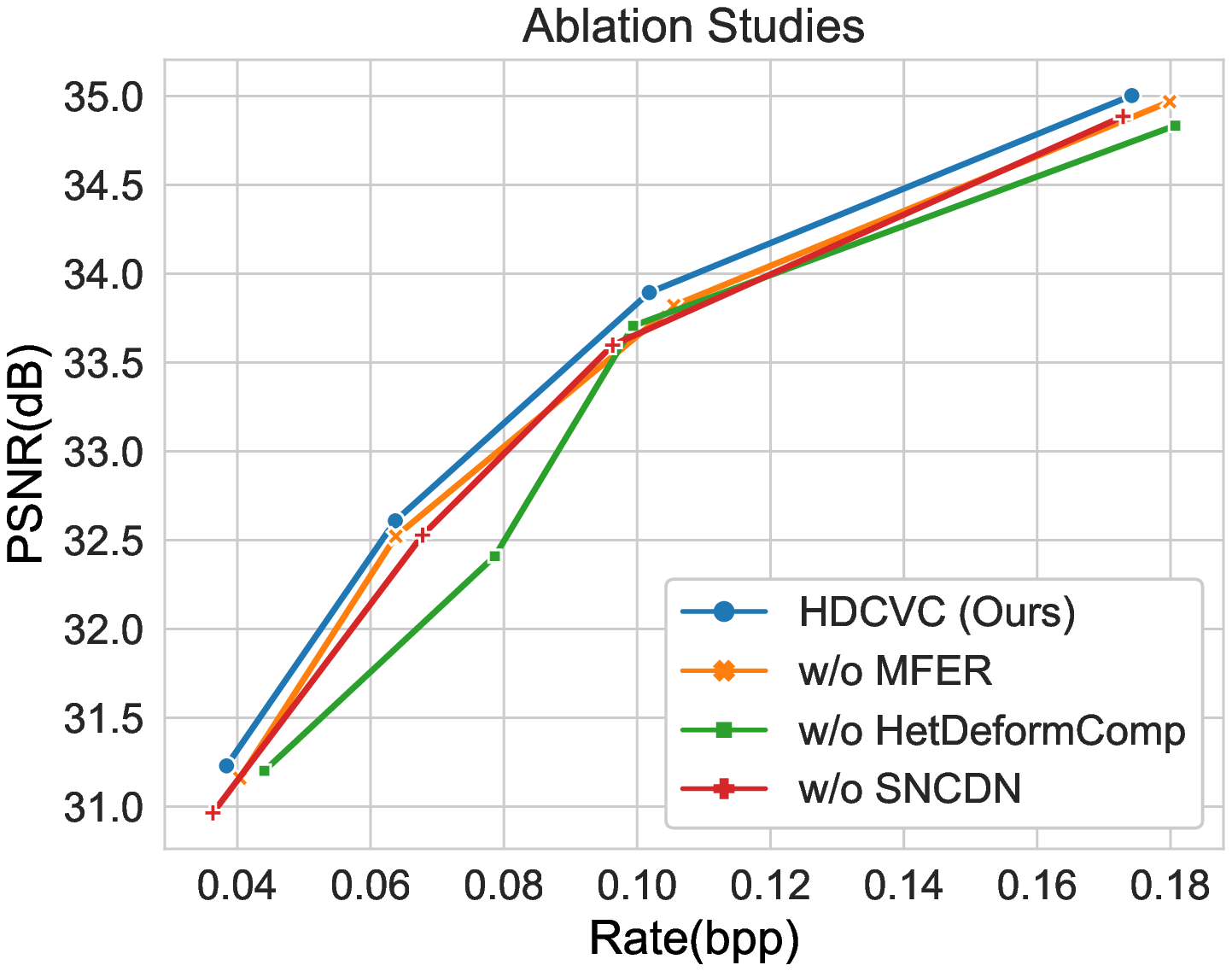}
		\caption{}
		\label{Ablation_PSNR}
	\end{subfigure}
	\begin{subfigure}{.32\textwidth}
		\centering
		\includegraphics[width=\textwidth]{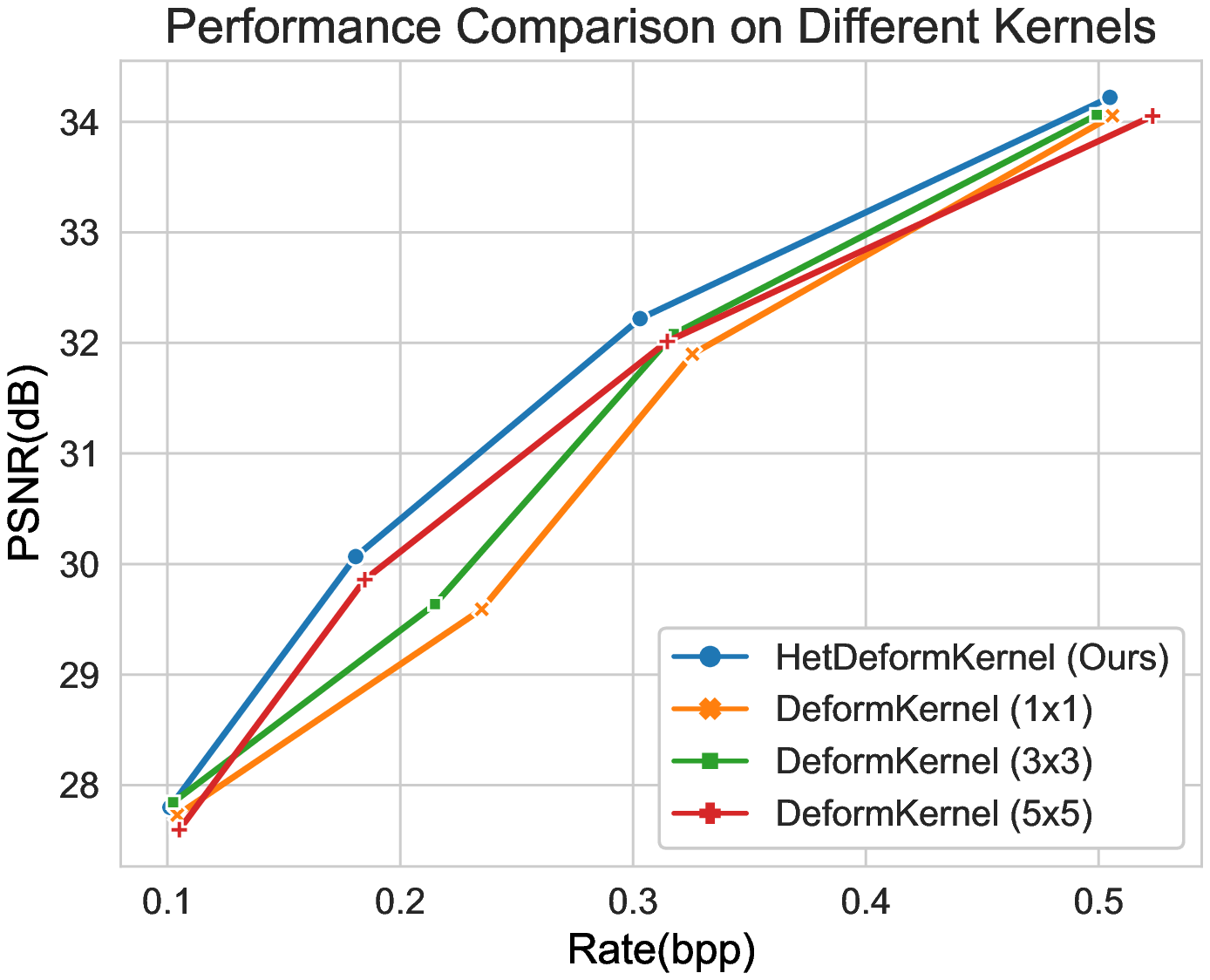}
		\caption{}
		\label{fig:comparison results}
	\end{subfigure}
	\begin{subfigure}{.32\textwidth}
		\centering
		\includegraphics[width=\textwidth]{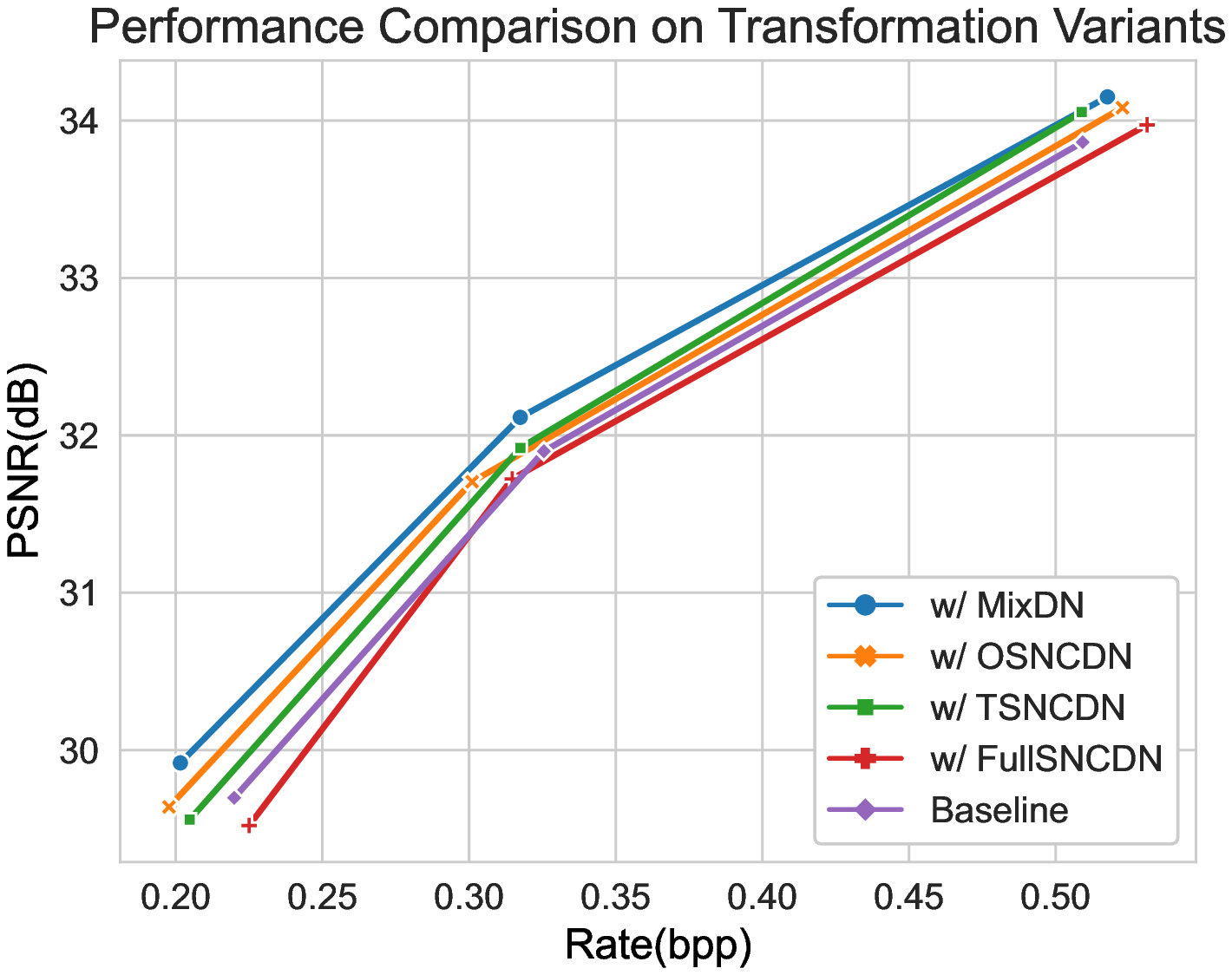}
		\caption{}
		\label{fig:SNCDN}
	\end{subfigure}
	\\
	\begin{subfigure}{.32\textwidth}
		\centering
		\includegraphics[width=\textwidth]{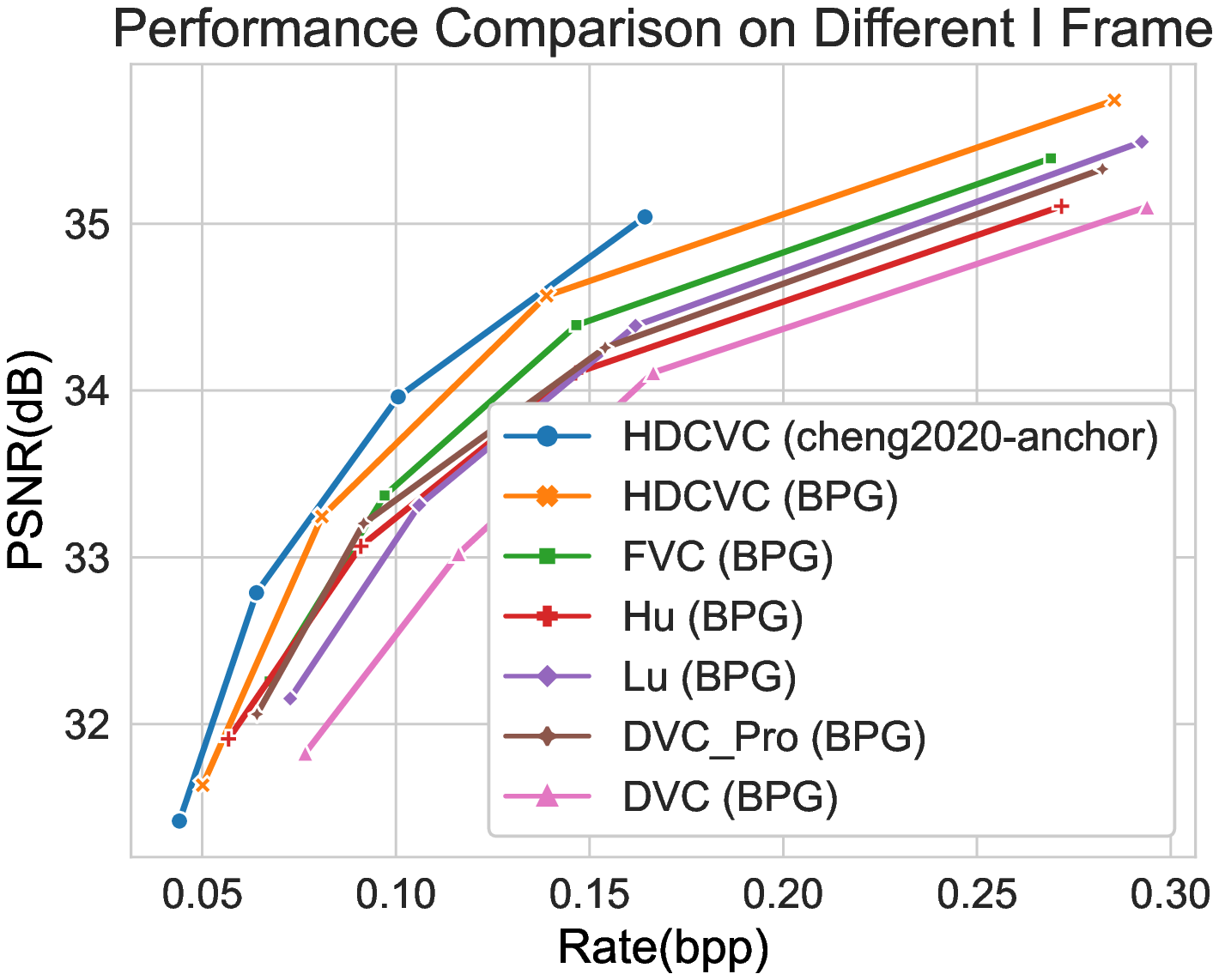}
		\caption{}
		\label{Ablation_I_Frame}
	\end{subfigure}
	\begin{subfigure}{.32\textwidth}
		\centering
		\includegraphics[width=\textwidth]{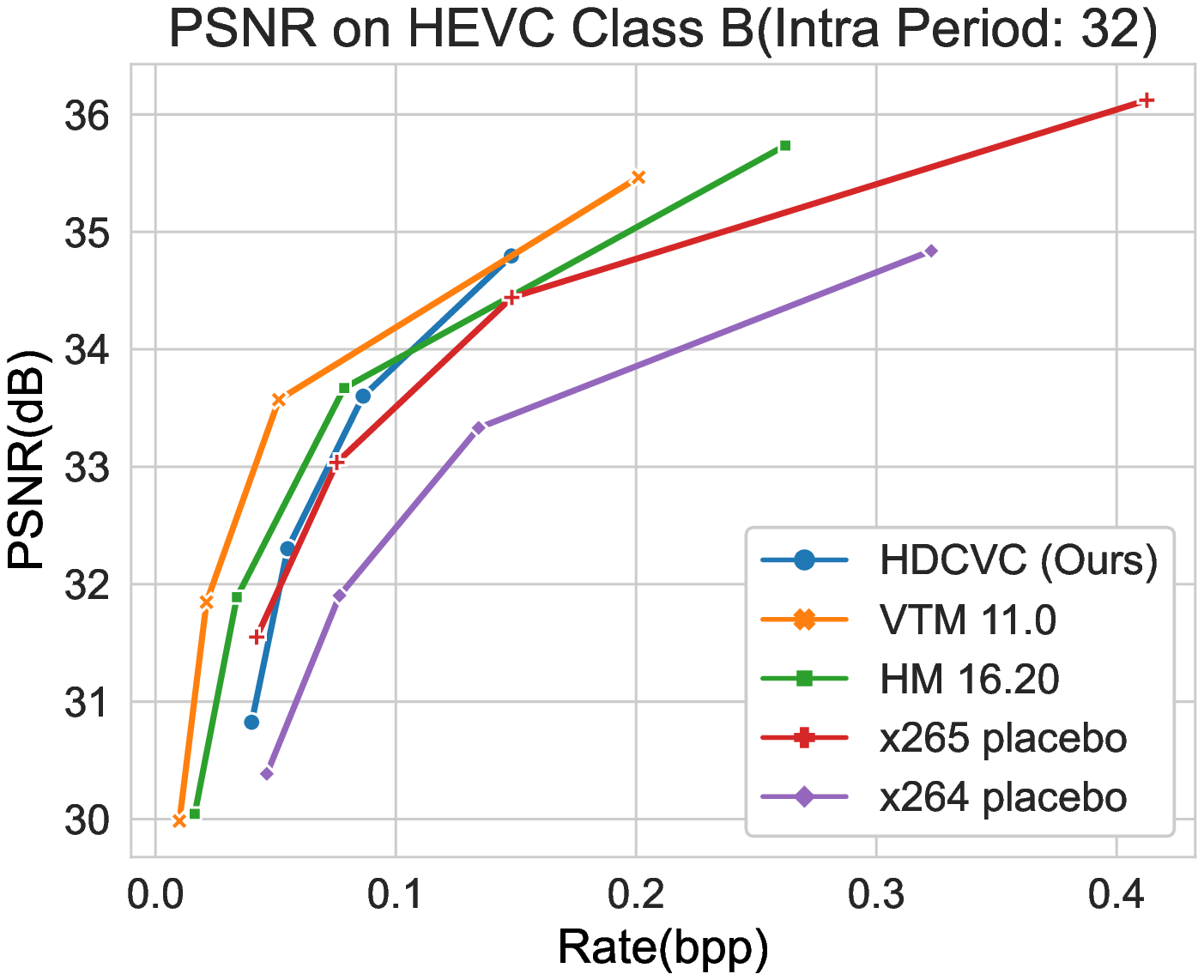}
		\caption{}
		\label{fig:comparison ClassB_IP32_PSNR}
	\end{subfigure}
	\begin{subfigure}{.32\textwidth}
		\centering
		\includegraphics[width=\textwidth]{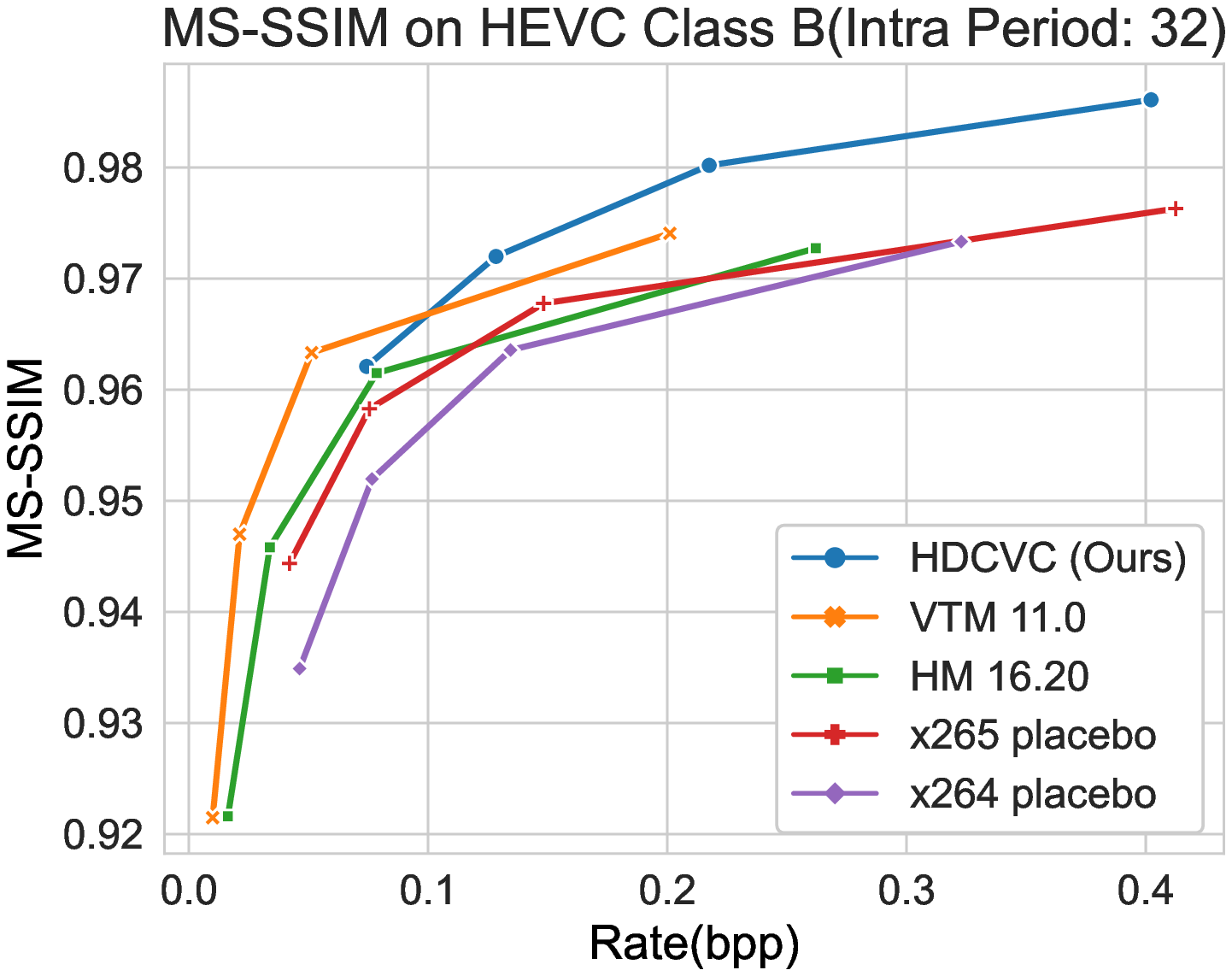}
		\caption{}
		\label{fig:ClassB_IP32_MSSSIM}
	\end{subfigure}
	\caption{Ablation studies and comparison with x264, x265, HM-16.20 and VTM-11.0 when intra period is set to 32.}
	\label{Ablation}
\end{figure*}

\begin{table}[t]
	\begin{center}
		\caption{Network inference time cost per frame of the HetDeform compensation and the single-size-kernel Deform compensation for a $1920\times1080$ sequence. All the comparison methods remove MFER module for a more direct comparison.}
		\label{time cost}
		\scalebox{1.2}{
			\begin{tabular}{c|c|c}
				\toprule[1.2pt]
				Methods              & Encoding time  & Decoding time   \\ \midrule
				HetDeform Baseline   & 273ms          & 350ms                 \\ \midrule
				$1\times 1$ Baseline & 246ms          & 368ms                 \\\midrule
				$3\times 3$ Baseline & 277ms          & 408ms                 \\\midrule
				$5\times 5$ Baseline & 415ms          & 561ms                 \\
				
				\bottomrule[1.2pt]
		\end{tabular}}
	\end{center}
\end{table}
\begin{figure*}[t]
	\centering
	\begin{subfigure}{.24\textwidth}
		\centering
		\includegraphics[width=\textwidth]{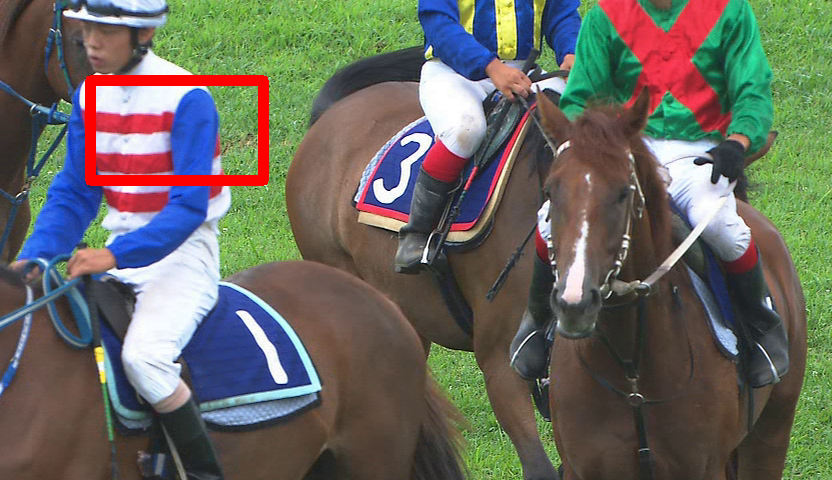}
		\caption{\textit{RaceHorses} }
		\label{fig:GT}
	\end{subfigure}
	\begin{subfigure}{.24\textwidth}
		\centering
		\includegraphics[width=\textwidth]{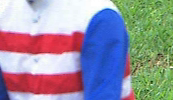}
		\caption{GT patch}
		\label{fig:croped_GT}
	\end{subfigure}  
	\begin{subfigure}{.24\textwidth}
		\centering
		\includegraphics[width=\textwidth]{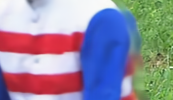}
		\caption{3x3 DeformCom}
		\label{fig:3x3kenel}
	\end{subfigure}
	\begin{subfigure}{.24\textwidth}
		\centering
		\includegraphics[width=\textwidth]{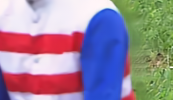}
		\caption{HetDeformCom}
		\label{fig:hetkenel}
	\end{subfigure}
	\caption{Visualization of the compensation results by single-size deformable convolution (DeformCom) and the proposed HetDeformConv for the 22nd frame of HEVC Class C \textit{RaceHorses}. The PSNR of the DeformCom and HetDeformCom are 25.76dB and 26.07db respectively, with similar bits cost for motion.}
	\label{Visualization}
\end{figure*}
\begin{figure*}[t!]
	\centering
	\begin{subfigure}{.31\textwidth}
		\centering
		\includegraphics[width=\linewidth]{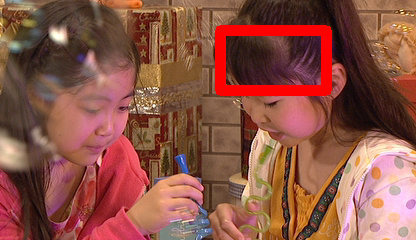}
		\caption{HEVC Class D BlowingBubbles}
	\end{subfigure}
	\begin{subfigure}{.315\textwidth}
		\centering
		\includegraphics[width=\linewidth]{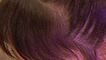}
		\caption{Ground Truth (MS-SSIM/BPP)}
	\end{subfigure}
	\begin{subfigure}{.315\textwidth}
		\centering
		\includegraphics[width=\linewidth]{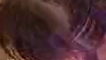}
		\caption{x264 placebo (0.9188/0.0812)}
	\end{subfigure}
	\\
	\begin{subfigure}{.315\textwidth}
		\centering
		\includegraphics[width=\linewidth]{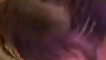}
		\caption{x265 default (0.9353/0.0881)}
	\end{subfigure}
	\begin{subfigure}{.315\textwidth}
		\centering
		\includegraphics[width=\linewidth]{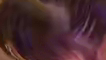}
		\caption{x265 placebo (0.9394/0.0899)}
	\end{subfigure}
	\begin{subfigure}{.315\textwidth}
		\centering
		\includegraphics[width=\linewidth]{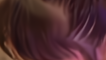}
		\caption{HDCVC (ours) \textbf{(0.9553/0.0779)}}
	\end{subfigure}
	\\
	\begin{subfigure}{.310\textwidth}
		\centering
		\includegraphics[width=\linewidth]{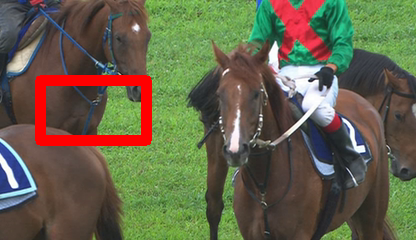}
		\caption{HEVC Class D RaceHorses}
	\end{subfigure}
	\begin{subfigure}{.315\textwidth}
		\centering
		\includegraphics[width=\linewidth]{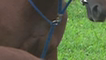}
		\caption{Ground Truth (MS-SSIM/BPP)}
	\end{subfigure}
	\begin{subfigure}{.315\textwidth}
		\centering
		\includegraphics[width=\linewidth]{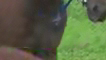}
		\caption{x264 placebo (0.9209/0.1212)}
	\end{subfigure}
	\\
	\begin{subfigure}{.315\textwidth}
		\centering
		\includegraphics[width=\linewidth]{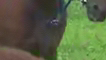}
		\caption{x265 default (0.9246/0.1252)}
	\end{subfigure}
	\begin{subfigure}{.315\textwidth}
		\centering
		\includegraphics[width=\linewidth]{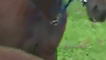}
		\caption{x265 placebo (0.9337/0.1277)}
	\end{subfigure}
	\begin{subfigure}{.315\textwidth}
		\centering
		\includegraphics[width=\linewidth]{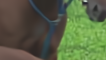}
		\caption{HDCVC (ours) \textbf{(0.9616/0.1161)}}
	\end{subfigure}
	\caption{The visual results of traditional hybrid codecs H.264, H.265 and our HDCVC (MS-SSIM model).}
	\label{MSSSIM_VisualResult}
\end{figure*}
In above settings, \textit{Num, Q, GOP} stand for the number of encoded frames, quality and intra period, respectively. In line with our test configuration, Num is set as 100 for the HEVC dataset. Following the configuration in \cite{9288876}, we set Q as 15, 19, 23, 27 for the HEVC and MCL-JCV datasets and 11, 15, 19, 23 for the UVG dataset. IP is set as 10 for the HEVC dataset and 12 for others.
	
We also provide the object results of x264, x265, HM-16.20 and VTM-11.0 when intra period is set to 32. For x264 and x265, it is simple to change intra period by changing \textit{keyint} parameter. As for HM-16.20 and VTM-11.0, we use the configuration files \textit{encoder\_lowdelay\_P\_main.cfg} and \textit{encoder\_lowdelay\_P\_vtm.cfg} for HM-16.20 and VTM-11.0, respectively. We use ffmpeg to convert the output yuv generated by the two methods losslessly into PNG format.

\begin{figure*}[t]
	\centering
	\begin{subfigure}{.32\textwidth}
		\centering
		\includegraphics[width=\linewidth]{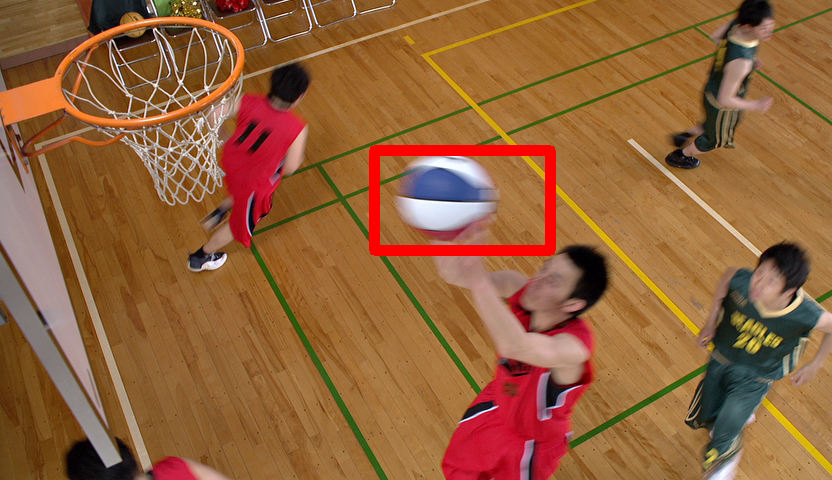}
		\label{BasketballDrive}
		\vspace{-12pt}
		\caption{HEVC Class B BasketballDrive}
	\end{subfigure}
	\begin{subfigure}{.325\textwidth}
		\centering
		\includegraphics[width=\linewidth]{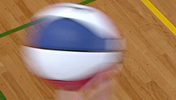}
		\label{BasketballDrive_GT}
		\vspace{-12pt}
		\caption{Ground Truth (PSNR (dB)/BPP)}
	\end{subfigure}
	\begin{subfigure}{.325\textwidth}
		\centering
		\includegraphics[width=\linewidth]{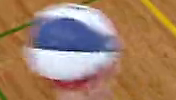}
		\label{BasketballDrive_x264_LDP_placebo}
		\vspace{-12pt}
		\caption{x264 placebo (28.29/0.0692)}
	\end{subfigure}
	\\
	\begin{subfigure}{.325\textwidth}
		\centering
		\includegraphics[width=\linewidth]{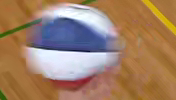}
		\label{BasketballDrive_x265_LDP_default}
		\vspace{-12pt}
		\caption{x265 default (30.056/0.0666)}
	\end{subfigure}
	\begin{subfigure}{.325\textwidth}
		\centering
		\includegraphics[width=\linewidth]{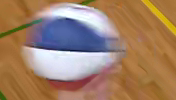}
		\label{BasketballDrive_x265_LDP_placebo}
		\vspace{-12pt}
		\caption{x265 placebo (30.16/0.0665)}
	\end{subfigure}
	\begin{subfigure}{.325\textwidth}
		\centering
		\includegraphics[width=\linewidth]{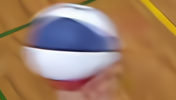}
		\label{HDCVC}
		\vspace{-12pt}
		\caption{HDCVC (ours) \textbf{(30.37/0.0618)}}
	\end{subfigure}
	\\
	\begin{subfigure}{.325\textwidth}
		\centering
		\includegraphics[width=\linewidth]{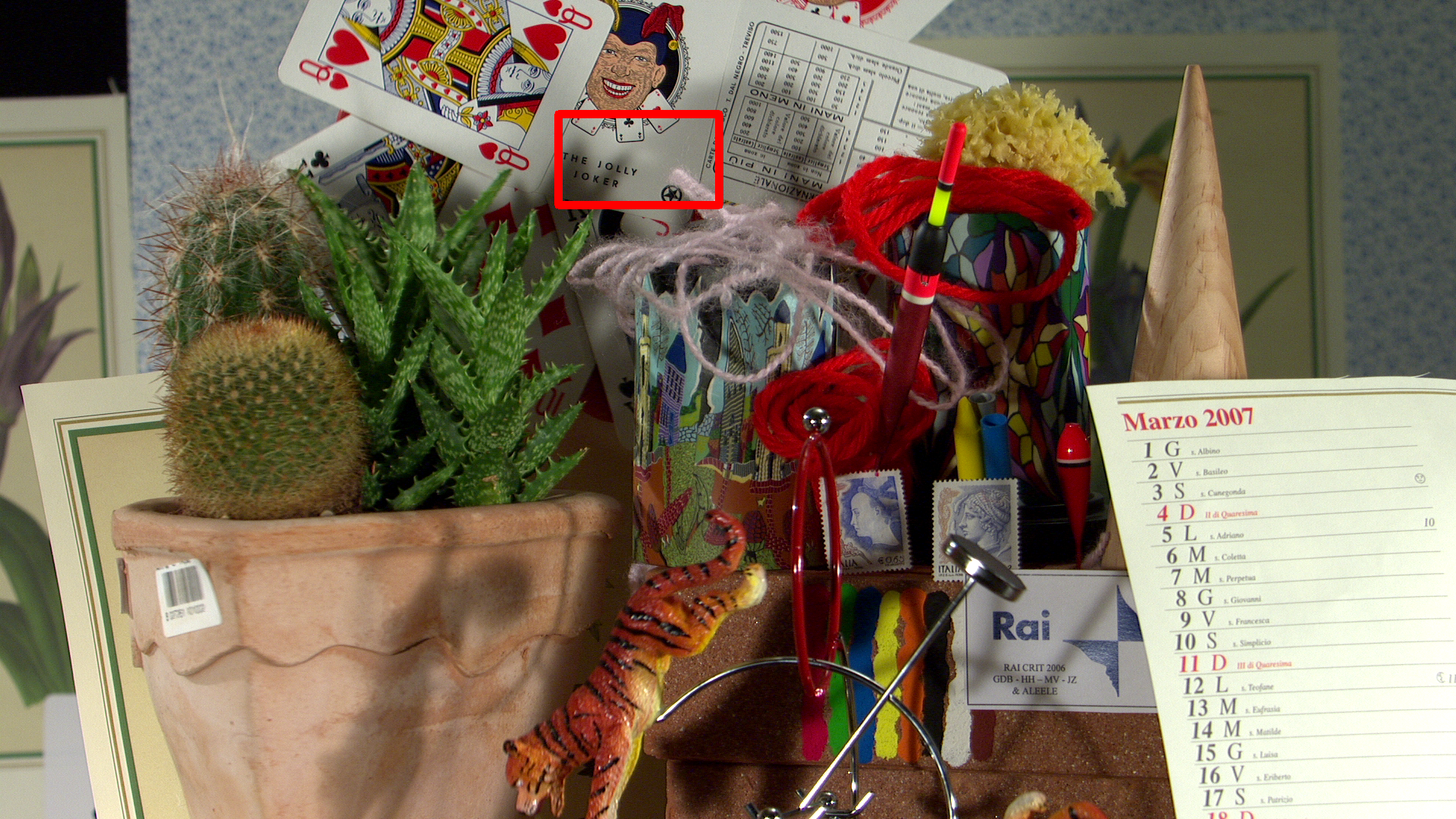}
		\label{Cactus}
		\vspace{-12pt}
		\caption{HEVC Class B  Cactus}
	\end{subfigure}
	\begin{subfigure}{.325\textwidth}
		\centering
		\includegraphics[width=\linewidth]{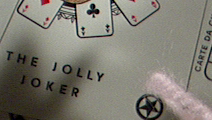}
		\label{Cactus_GT}
		\vspace{-12pt}
		\caption{Ground Truth (PSNR (dB)/BPP)}
	\end{subfigure}
	\begin{subfigure}{.325\textwidth}
		\centering
		\includegraphics[width=\linewidth]{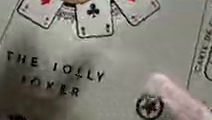}
		\label{Cactus_x264_LDP_placebo}
		\vspace{-12pt}
		\caption{x264 placebo (28.88/0.0485)}
	\end{subfigure}
	\\
	\begin{subfigure}{.325\textwidth}
		\centering
		\includegraphics[width=\linewidth]{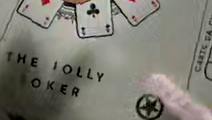}
		\label{Cactus_x265_LDP_default}
		\vspace{-12pt}
		\caption{x265 default (30.33/0.0467)}
	\end{subfigure}
	\begin{subfigure}{.325\textwidth}
		\centering
		\includegraphics[width=\linewidth]{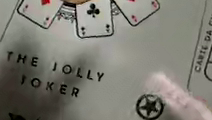}
		\label{Cactus_x265_LDP_placebo}
		\vspace{-12pt}
		\caption{x265 placebo (30.50/0.0457)}
	\end{subfigure}
	\begin{subfigure}{.325\textwidth}
		\centering
		\includegraphics[width=\linewidth]{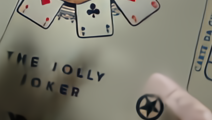}
		\label{Cactus_HDCVC}
		\vspace{-12pt}
		\caption{HDCVC (ours) \textbf{(30.68/0.0425)}}
	\end{subfigure}
	\caption{The visual results of traditional hybrid codecs H.264, H.265 and our HDCVC (PSNR model).}
	\label{PSNR_VisualResult}
\end{figure*}

\subsection{Comparison with Other Methods}
We compare our method with recent video compression methods: DVC\cite{Lu2019CVPR}, DVC\_Pro\cite{lu2020end}, HLVC\cite{yang2020learning}, SSF\cite{agustsson2020scale}, Hu\cite{hu2020improving}, Liu2022\cite{9707786} and FVC\cite{hu2021fvc}. As for the traditional method H.265\cite{HEVC}, we refer to the comand line in \cite{Lu2019CVPR,hu2020improving} and use x265\cite{x265} with the LDP (Low Delay P) placebo mode. To be consistent with the test condition on \cite{Lu2019CVPR,yang2020learning,hu2020improving}, we test HEVC on the first 100 frames and test UVG, MCL-JCV sequences on all frames, and intra period is set to 10 for the JCT-VC dataset and 12 for other datasets. To compare with HM-16.20 and VTM-11.0 in more reasonable conditions, we also provide the RD results of HDCVC when intra period is set to 32 ( see Fig.~\ref{fig:comparison ClassB_IP32_PSNR} and \ref{fig:ClassB_IP32_MSSSIM}).

\paragraph{Rate-distortion performance.}
Fig.~\ref{Result} shows the experimental results on the HEVC, UVG and MCL-JCV sequences while taking PSNR and MS-SSIM as quality measurements. The RD performance of our PSNR model can surpasses other learned methods and H.265 (x265 LDP placebo) significantly. As for the results measured by MS-SSIM, our model still achieves comparable performance among recent video compression approaches on all test datasets.

\paragraph{BDBR results.}
To show quantitative performance comparisons with current advanced methods, we provide Bj\o ntegaard Delta Bit-Rate (BDBR)\cite{bjontegaard2001calculation} with the anchor of x265 (LDP placebo). BDBR represents the calculation of average PSNR or other evaluation metrics differences between RD-curves, and it also denotes the bits saving at the same quality compared to the anchor. DVC\cite{Lu2019CVPR}, DVC\_Pro\cite{lu2020end}, HLVC\cite{yang2020learning}, Lu\cite{lu2020content}, Hu\cite{hu2020improving} and FVC\cite{hu2021fvc} are compared with HDCVC on the HEVC, UVG and MCL-JCV datasets.

With the merit of the proposed schemes, our PSNR model shows competitive RD results with other learned video compression methods. Our MS-SSIM model also surpasses other learning-based approaches on most datasets. Table~\ref{BDBR} shows the detailed BDBR results calculated by PSNR and MS-SSIM.

\subsection{Ablation Studies and Performance Analysis}
\subsubsection{HetDeform Compensation}\label{sec:ablation_HetDeform}
To verify the effectiveness of the HetDeform Compensation, we conduct an experiment by replacing the HetDeform compensation in our model with standard $3\times3$ kernel deformable compensation on HEVC Class B (denoted by w/o HetDeformComp, the orange curve in Fig.~\ref{Ablation_PSNR}). In the absence of the introduced compensation method, The PSNR will decrease by 0.4 to 0.8 dB at the same bpp.

We also conduct experiments by changing the kernel size of the deformable compensation module in our Baseline (HDCVC w/o HetDeformConv and SNCDN) model. As shown in Fig.~\ref{fig:comparison results}, deformable-compensation-based methods cannot achieve globally optimal performance. The visualization comparison in Fig.~\ref{Visualization} can further demonstrate the superiority of our compensation scheme. Visually, it can be observed that the compensation results of our scheme achieve better quality and can reduce obvious compensation inaccuracies.

\subsubsection{SNCDN}
To evaluate the effectiveness of our proposed normalization method, we replace the SNCDN and MixDN in the transformation network by GDN (denoted by w/o SNCDN, the green curve in Fig.~\ref{Ablation_PSNR}). When compared with our complete model, we can observe that using the SNCDN and MixDN provides nearly 0.25 dB in PSNR at the same bit rate.

To further explore the effectiveness of the proposed SNCDN, we conduct a series of experiments on variants of the transformation network. w/ MixDN (the blue curve in Fig.~\ref{fig:SNCDN}) is our default arrangement as Fig.~\ref{sec:encoder-decoder} shows. We replace the first one, first two, and all ResGDN in the baseline with SNCDN (denoted by w/ OSNCDN, w/ TSNCDN, and w/ FullSNCDN). From the comparison result, we infer that SNCDN can play a better transformation effect when the feature is at a higher resolution, and a suitable location will yield higher performance.

\subsubsection{MFER}
To demonstrate the effectiveness of the proposed MFER scheme, we compare the compression performance of the HDCVC with or without using the reconstruction module. In the absence of this module, the PSNR will decrease by 0.2dB at the same bpp. Besides, with MFER, HDCVC can surpass HM-16.20 and VTM-11.0 in terms of MS-SSIM in the high bit rate range by a large margin when intra period is set to 32, as shown in Fig.~\ref{fig:ClassB_IP32_MSSSIM}.

\subsubsection{Different I frame}
In order to compare with other learning-based methods in a more similar situation, we conduct extra experiments with BPG as the I frame compression method. As shown in Fig.~\ref{Ablation_I_Frame}, without cheng2020-anchor as I frame compression, HDCVC still achieves the highest RD performance when compared with other SOTA methods.

\paragraph{Model complexity and running time.}
Our proposed model have about 32M parameters, while the comparison algorithm FVC\cite{hu2021fvc} has 26M parameters. We take the sequences with the resolution of $1920 \times 1080$ to test inference time. We only record network inference time of coding for a more precise time-consuming comparison of our proposed modules. We denote HetDeform compensation baseline without MFER as HetDeform Baseline, $1\times 1$ deformable compensation baseline without MFER as $1\times 1$ Baseline and so on. The network encoding/decoding time can refer to Table~\ref{time cost}. We can observe from the results that the proposed compensation scheme combines the advantages of high efficiency and high performance through a reliable implementation.

\subsection{Visualization results}
This section presents the visual quality comparison of H.264, H.265 and our proposed models. We select four representative sequences in the HEVC dataset, including \textit{BasketballDrive and Cactus}. Fig.~\ref{PSNR_VisualResult} and Fig.~\ref{MSSSIM_VisualResult} shows the visual quality comparison of the H.264 (x264 LDP placebo), H.265 (x265 LDP default), H.265 (x265 LDP placebo) and our HDCVC. With fewer bits consuming, the proposed model produces fewer compression artifacts and achieves higher subjective quality and less color shift than the traditional hybrid codecs H.264 and H.265.

\section{Conclusion}
In this paper, we present a learned video compression scheme via a novel heterogeneous deformable compensation network HDCVC. Specifically, the proposed model can generate content-adaptive offsets and apply the heterogeneous deformable convolution to accomplish motion compensation. The proposed approach addresses the problem of non-globally optimal performance caused by single-size-kernel deformable compensation in downsampled feature domain. The SNCDN is designed to assist in compressing motion and residual into latent representations, using a set of strongly correlated neighborhoods for Gaussianization and statistic dependencies reduction. Furthermore, the MFER module is also proposed for exploiting context and temporal information from decoded features and frames, introducing long-range temporal information to alleviate error propagation. Comprehensive experiments have been performed on several test datasets, and the experimental results can demonstrate the effectiveness of our proposed framework.

%%%%%%%%% REFERENCES
\bibliographystyle{IEEEtran}
\bibliography{egbib}
\leavevmode
\newline
\textbf{Huairui Wang} is pursuing the Ph.D. degree at Wuhan University, China.
\\
\\
\textbf{Zhenzhong Chen} is a Professor at Wuhan University, China.
\\
\\
\textbf{Chang Wen Chen} is Chair Professor of Visual Computing at The Hong Kong Polytechnic University, China.

\end{document}